\documentclass[11pt,draftclsnofoot,onecolumn]{IEEEtran}

\usepackage[dvips]{graphicx}
\usepackage{color}
\usepackage{bm}
\usepackage{psfrag}
\usepackage[mathscr]{eucal}
\usepackage{times}
\usepackage{epsfig,graphicx,color,pstricks}
\usepackage{pst-node,pst-blur}
\usepackage{amsmath,amssymb,amsfonts,eucal,latexsym,mathcom}
\usepackage{cite}

\begin{document}
\IEEEpeerreviewmaketitle
\newtheorem{thm}{theorem}

\title{Outage analysis of Block-Fading Gaussian Interference Channels}


\author{
\authorblockN{Yang~Weng and Daniela~Tuninetti\\}
\authorblockA{University of Illinois at Chicago, IL (USA),
              Department of Electrical and Computer Engineering,
              Email: {\tt yweng3@uic.edu, danielat@uic.edu}.} %
}
\maketitle

\begin{abstract}
This paper considers the asymptotic behavior of two-source block-fading
single-antenna Gaussian interference channels in the high-SNR regime
by means of the diversity-multiplexing tradeoff. We consider a general
setting where the users and the average channel gains are not restricted
to be symmetric. Our results are not just extensions of previous
results for symmetric networks, as our setting covers scenarios that
are not possible under the symmetric assumption, such as the case of
``mixed'' interference, i.e., when difference sources have different
distances from their intended receivers.
We derive upper and lower bounds on the diversity.
We show that for a fairly large set of channel parameters
the two bounds coincides.
\end{abstract}

\section{Introduction}
\label{sec:intro}
Wireless networks deal with two fundamental limits
that make the communication problem challenging and interesting.
On the one hand, simultaneous communications from uncoordinated
users create undesired interference. On the other hand, fluctuations
of the channel condition due to multi-path and mobility cause
signals to fade randomly.
In today's cellular and ad-hoc networks
orthogonalization techniques, such as F/T/C/SDMA, are employed to
avoid interference.  However, although leading to simple network
architectures, interference avoidance techniques are suboptimal in
terms of achievable rates.
Moreover, the relative strength of the intended data signal
and the interference signals changes over time due to fading.
This makes fixed channel access strategies suboptimal.
Thus, understanding how to deal simultaneously with
interference and with fading holds the key to the deployment
of future broadband wireless networks.
The simplest model for analyzing these problems jointly
is the two-source Block-Fading Gaussian InterFerence Channel
(BF-GIFC).

It is well know that the Han-Kobayashi
(HK)~\cite{Han_Kobayashi:it1981} scheme with superposition coding,
rate splitting, and joint decoding, gives the largest known achievable
rate region  for GIFC without fading.
Several outer bounds are known in the literature for
GIFC without fading~\cite{kramer:ifcout:it04,sason:ifc:it04,etkin_tse_hua:withinonebit:subIt06,
shang:ifcout:it09,Annapureddy:ifout:it09,Motahari:ifcout:it09}.
In particular, Etkin et al.~\cite{etkin_tse_hua:withinonebit:subIt06} showed that a
simple rate splitting strategy in the HK scheme is within one~bit/sec/Hz of the
capacity region of Gaussian unfaded GIFCs for any possible channel parameters.
In~\cite{etkin_tse_hua:withinonebit:subIt06},
all interfering signals above the noise floor are decoded, that is,
the private messages --which are treated as noise-- are assigned a
transmit power such that they are going to be received at, or below, the level of the
noise.  In doing so, roughly speaking, the effective noise power at the receiver is
at most doubled, thus giving a rate penalty of at most $1$~bit/sec/Hz.

Recently, GIFCs with fading were considered in~\cite{Cadambe-Jafar:arXiv:0802.2125,
Sankar:isit08, tuninetti:asilomar2008, leveque:izs2008,AkcabaBolcskeiisit09,
RajaViswanathisit09, EbrahimzaKhandaniisit09, SezginJafarJaf}.

For ergodic channels, such as fast fading channels, the (Shannon) capacity
is the performance measure of the ultimate system performance.
In~\cite{Cadambe-Jafar:arXiv:0707.0323v2}, it was showed that the sum-rate ergodic
capacity of a $K$-source fading GIFC scales linearly with the number of sources.
In~\cite{Sankar:isit08}, the sum-rate capacity of a two-source strong ergodic fading
GIFCs was shown to be equal to that of the corresponding compound MAC.
In~\cite{tuninetti:asilomar2008}, optimal power allocation policies
for outer and inner bounds for ergodic fading GIFCs with perfect transmitter
CSI were derived.

For slow fading channels, the proper measurement of
performance is the outage capacity. In particular, the Diversity Multiplex Tradeoff
(DMT)~\cite{zheng-tse:dmt},
quantifies the tradeoff between rate and outage probability
as the Signal to Noise Ratio (SNR) grows to infinity.
In~\cite{leveque:izs2008} the DMT of symmetric two-source BF-GIFCs
was studied based on the ``within one bit'' outer bound of~\cite{etkin_tse_hua:withinonebit:subIt06}.
The authors of~\cite{leveque:izs2008} claimed that the derived DMT
is actually achievable because the ``one~bit penalty'' for using a simple
HK strategy vanishes at high SNR. However, the achievability of the ``within one bit''
outer bound requires a very specific rate splitting in the HK achievable scheme
that depends on the instantaneous fading values.  Hence, as pointed out in~\cite{YangTuninetti:spac2009,YangTuninetti:allerton2009,RajaViswanathisit09,
AkcabaBolcskeiisit09,EbrahimzaKhandaniisit09}
the DMT derived in~\cite{leveque:izs2008} is achievable only if the transmitters know
the instantaneous fading values perfectly. In the the case of no channel state
information at the transmitter (TXCSI)
the DMT of~\cite{leveque:izs2008} is an upper-bound on the actual DMT.

The DMT of BF-GIFCs without TXCSI is the subject of investigation of~\cite{YangTuninetti:spac2009,YangTuninetti:allerton2009,RajaViswanathisit09,
AkcabaBolcskeiisit09,EbrahimzaKhandaniisit09,SezginJafarJaf} as well as of this work.
In~\cite{AkcabaBolcskeiisit09}, it was proved that in strong interference
joint decoding of all message at all destinations achieves the DMT outer-bound
of~\cite{leveque:izs2008}.
In~\cite{RajaViswanathisit09}, it was showed that multilevel superposition coding
achieves the DMT of any two-source BF-GIFC; however, no explicit formula was given
for more than two levels of superposition.
In~\cite{SezginJafarJaf}, it was showed that the DMT of BF-GIFCs reduces
to that of Multiple Access Channel (MAC) if transmitters are not aware of the channel gains.
In~\cite{EbrahimzaKhandaniisit09}, it was shown that one~bit of TXCSI suffices to achieve
the optimum DMT for certain rages of channel parameters.

The works~\cite{leveque:izs2008,AkcabaBolcskeiisit09,EbrahimzaKhandaniisit09,SezginJafarJaf}
focused on two-source symmetric networks, that is to say, networks
for which the average SNR and the average Interference to Noise Ratio (INR) at all
receivers are the same. In this work, we consider two-source {\em asymmetric}
GIFCs as in our conference papers~\cite{YangTuninetti:spac2009,YangTuninetti:allerton2009,RajaViswanathisit09}.
In~\cite{YangTuninetti:spac2009,YangTuninetti:allerton2009} we generalized the DMT
outer-bound of~\cite{leveque:izs2008} to asymmetric networks and studied
HK achievable schemes with and without rate splitting.
It should be point out that our results are not just a generalization
of the symmetric network results.  Our setting covers all possible classes of channels
and includes channels not possible under the symmetric assumptions,
such as the case of ``mixed'' interference.  Mixed interference
occurs in practice when
sources have different distances from their intended receivers and is
the most practical scenario for wireless networks.


We assume that the channel variations
over time are sufficiently slow so that the fading coefficients may
be considered as fixed for the whole codeword duration (i.e., block
fading assumption). We assume that the receivers know perfectly the
channel realization, but the transmitters do not.  In this case, if
the instantaneous fading realization is such that the transmission
rates cannot be reliably decoded, the system is said to experience
{\em outage}. In an outage setting without TXCSI, it is
not clear that a fixed rate splitting strategy can actually achieve
the DMT upper bound of~\cite{leveque:izs2008}. Here we consider both HK
achievable strategies with and without rate splitting.  In the
case of rate splitting, we consider the case where the average received
power of the signals that are treated as noise is set below the noise floor.
as done in~\cite{etkin_tse_hua:withinonebit:subIt06} for the unfaded case.
We also generalized the outer-bound of~\cite{leveque:izs2008} to asymmetric
networks. We show that for a very wide range of channel parameters, the
inner and outer bound meet. In particular, rate splitting improves the
achievable DMT in weak and mixed interference channel.

The rest of the paper is organized as follows:
Section~\ref{sec:chmodel} presents the system model and the problem formulation;
Section~\ref{sec:upperbound} and~\ref{sec:lowerbound} present DMT upper and lower
bounds, respecetively;
Section~\ref{sec:num} presents numerical results;
Section~\ref{sec:con} concludes the paper.

\section{Channel Model}
\label{sec:chmodel}

A two-source single-antenna Rayleigh fading GIFC
in standard form is defined as:
\begin{align}
Y_u &= H_{u1}X_1+H_{u2}X_2+Z_u \in \CC,
\label{eq:channelmodel}
\end{align}
where the noises $Z_u\sim \mathcal{N}(0, 1)$ and the inputs
are subject to the average power constraint
$ \E[|X_u|^2] \leq P_u$, $u\in\{1, 2\}$.
We assume the channel to be block-fading and that each codeword
spans one fading block, i.e., no coding across multiple blocks is allowed.
Moreover, we assume arbitrarily large block lengths.
The receivers are assumed to perfectly know the fading realization
$(H_{11},H_{12},H_{21},H_{22})$, while the transmitters are not. In
the rest of the paper we parameterize the received SNR/INRs as
\begin{align}
\E[|H_{cu} X_{u}|^2] =
\E[|H_{cu}|^2]P_u\defeq  x^{\beta_{cu}},
\quad \beta_{cu}\in\RR^+,
\quad (c,u)\in\{1, 2\}\times\{1, 2\},
\label{eq:snr/inr parametrization}
\end{align}
for some $x>1$, and the transmission rates as
\begin{align}
R_u \defeq \log(1+x^{r_u}),
\quad r_{u}\in\RR^+,
\quad u \in\{1, 2\}.
\label{eq:rate parametrization}
\end{align}
We focus our analysis on the high-SNR regime,
that is, in the limit for $x\to+\infty$. Notice that, although
we impose that the channel gains $\beta$'s and the rates
$r$'s to be non-negative, the results derived in the following
can be extended to any $\beta$'s and $r$'s by replacing
each $\beta$ with $[\beta]^+\defeq\max\{0,\beta\}$ and
each $r$ with $[r]^+\defeq\max\{0,r\}$.

\paragraph{Capacity outer bound}
The capacity region of GIFC is not known in general.
Recently, Etkin et al.~\cite{etkin_tse_hua:withinonebit:subIt06}
proposed a novel outer bound for the capacity region of unfaded GIFC
that is shown to be ``within one bit'' of a simplified version of
the Han-Kobahyashy~\cite{Han_Kobayashi:it1981} achievable region.
More precisely, let
\begin{align}
H_{ij} = \sqrt{\E[|H_{ij}|^2]\,x^{-\gamma_{ij}}} \, \eu^{\jj \theta_{ij}}
\label{eq:fading parametrization}
\end{align}
where $(\gamma_{ij},\theta_{ij})$ are iid for all $(i,j)$ and with
$\theta_{ij}$ uniformly distributed on $[0,2\pi]$ and independent
of $\gamma_{ij}$.
In~\cite{zheng-tse:dmt} it is shown that in the limit for
$x\to+\infty$, the random variables $\gamma_{ij}$ are asymptotically
iid negative exponential with mean $\log(x)$. By using the
parameterization~\reff{eq:snr/inr parametrization},~\reff{eq:rate parametrization},
and~\reff{eq:fading parametrization}, for each fading realization
$(\gamma_{11},\gamma_{12},\gamma_{21},\gamma_{22})$, the ``within
one bit'' outer bound of~\cite{etkin_tse_hua:withinonebit:subIt06}
can be written as~\cite{leveque:izs2008}:
\begin{subequations}
\begin{align}
&\mathcal{R}_{\rm ETW}=\bigg\{
(\gamma_{11},\gamma_{12},\gamma_{21},\gamma_{22})\in\RR^4: \nonumber
\\&\log(1+x^{r_1})\leq \log(1+x^{\beta_{11}-\gamma_{11}})\label{eq:a}
\\&\log(1+x^{r_2})\leq \log(1+x^{\beta_{22}-\gamma_{22}})\label{eq:b}
\\&\log(1+x^{r_1})+\log(1+x^{r_2}) \leq \nonumber
\log(1+\frac{x^{\beta_{11}-\gamma_{11}}}{1+x^{\beta_{21}-\gamma_{21}}})\\&\qquad
~~~+\log(1+x^{\beta_{22}-\gamma_{22}}+x^{\beta_{21}-\gamma_{21}})\label{eq:c}
\\&\log(1+x^{r_1})+\log(1+x^{r_2}) \leq \nonumber
\log(1+\frac{x^{\beta_{22}-\gamma_{22}}}{1+x^{\beta_{12}-\gamma_{12}}})\\&\qquad
~~~+\log(1+x^{\beta_{11}-\gamma_{11}}+x^{\beta_{12}-\gamma_{12}})\label{eq:d}
\\&\log(1+x^{r_1})+\log(1+x^{r_2})\leq \nonumber
\\&\qquad ~~~
\log(1+x^{\beta_{12}-\gamma_{12}}+\frac{x^{\beta_{11}-\gamma_{11}}}{1+x^{\beta_{21}-\gamma_{21}}})\nonumber
\\&\qquad+
\log(1+x^{\beta_{21}-\gamma_{21}}+\frac{x^{\beta_{22}-\gamma_{22}}}{1+x^{\beta_{12}-\gamma_{12}}})\label{eq:e}
\\
&2\log(1+x^{r_1})+\log(1+x^{r_2})\leq
\log(1+\frac{x^{\beta_{11}-\gamma_{11}}}{1+x^{\beta_{21}-\gamma_{21}}})\nonumber
\\&\qquad+
\log(1+x^{\beta_{11}-\gamma_{11}}+x^{\beta_{12}-\gamma_{12}})\nonumber
\\& \qquad+
\log(1+x^{\beta_{21}-\gamma_{21}}+\frac{x^{\beta_{22}-\gamma_{22}}}{1+x^{\beta_{12}-\gamma_{12}}})\label{eq:f}
\\&\log(1+x^{r_1})+2\log(1+x^{r_2})\leq
\log(1+\frac{x^{\beta_{22}-\gamma_{22}}}{1+x^{\beta_{12}-\gamma_{12}}})
\nonumber
\\&\qquad+
\log(1+x^{\beta_{22}-\gamma_{22}}+x^{\beta_{21}-\gamma_{21}}) \nonumber
\\&\qquad+
\log(1+x^{\beta_{12}-\gamma_{12}}+\frac{x^{\beta_{11}-\gamma_{11}}}{1+x^{\beta_{21}-\gamma_{21}}})
\label{eq:g} \bigg\}.
\end{align}
\label{eq:outerbound within one bit}
\end{subequations}

\subsection{Capacity inner bound}
The HK achievable
region, in a form that matches the rate bounds
of~\reff{eq:outerbound within one bit}, can be found
in~\cite{Fah_Garg_Motani:sub2006}, and is given by:
\begin{align*}
\mathcal{R}_{\rm HK,complete}&=
\bigcup_{P(Q,W_1,W_2,X_1,X_2)}\bigg\{
(R_{1},R_{2})\in\RR^2_+: \nonumber\\
R_1&\leq I(X_1;Y_1|W_2Q); \\
R_2&\leq I(X_2;Y_2|W_1Q); \\
R_1+R_2 &\leq I(X_2,W_1;Y_2|Q)    +I(X_1;Y_1|W_1W_2,Q); \\
R_1+R_2 &\leq I(X_1,W_2;Y_1|Q)    +I(X_2;Y_2|W_1W_2,Q); \\
R_1+R_2 &\leq I(X_1,W_2;Y_1|W_1,Q)+I(X_2,W_1;Y_2|W_2,Q); \\
2R_1+R_2&\leq I(X_1,W_2;Y_1|Q)    +I(X_1;Y_1|W_1,W_2,Q)+I(X_2,W_1;Y_2|W_2,Q)\\
R_1+2R_2&\leq I(X_2,W_1;Y_2|Q)    +I(X_2;Y_2|W_1,W_2,Q)+I(X_1,W_2;Y_1|W_1,Q) \bigg\}.
\end{align*}

The region $\mathcal{R}_{\rm HK,complete}$ is difficult to
evaluate because it requires an optimization with respect to the
joint distribution $P(Q,W_1,W_2,X_1,X_2)$, where $Q$ is a
time-sharing random variable and $(W_1,W_2)$ has the meaning of
common information decoded at both receivers.  In order to have a
region that can be evaluated easily, it is customary to assume
jointly Gaussian input $P(W_1,W_2,X_1,X_2|Q)$
without time sharing, that is, the random variable $Q$ is a deterministic constant. We
set $W_u\sim \Nc(0,P_{u,common})$ independent of $T_u\sim
\Nc(0,P_{u,private})$ and let $X_u=W_u+T_u$ such that the total power
constraint is met with equality, i.e., $P_u =P_{u,private}+P_{u,common}$, for
$u\in\{1,2\}$.  We further parameterize the ratio of the average private
power to the total average power for a given user $u$ as
\begin{align}
\alpha_u
= \frac{1}{1+x^{b_u}}\in[0,1], \quad b_u\in\RR,
\end{align}
so that the average receive SNR/INR's on channel $c\in\{1,2\}$ are
\begin{align*}
\E[|H_{cu}|^2]P_{u,private}&=   \alpha_u \, \E[|H_{cu}|^2]P_u = \frac{x^{\beta_{cu}}}{1+x^{b_u}}
      \\
\E[|H_{cu}|^2]P_{u,common}&=(1-\alpha_u)\, \E[|H_{cu}|^2]P_u = \frac{x^{\beta_{cu}+b_u}}{1+x^{b_u}},
\quad u\in\{1,2\}.\\
\end{align*}
Moreover, following~\cite{etkin_tse_hua:withinonebit:subIt06},
we set  $b_1=\beta_{21}$ and $b_2=\beta_{12}$ so that
the average interfering private
power is below the level of the noise,
that is,
$\E[|H_{cu}|^2]P_{u,private}= \frac{x^{b_{u}}}{1+x^{b_u}} \leq 1$.

With these choices, $\mathcal{R}_{\rm HK,complete}$ reduces to
\begin{subequations}
\begin{align}
\mathcal{R}_{\rm HK}=
\bigg\{&
(\gamma_{11},\gamma_{12},\gamma_{21},\gamma_{22})\in\RR^4: \nonumber
\\&\log(1+x^{r_1}) \leq
   \log\left(1+\frac{x^{\beta_{11}-\gamma_{11}}}{1+\frac{x^{\beta_{12}-\gamma_{12}}}{1+x^{+\beta_{12}}}}\right)
   \label{eq:HK a}
\\&\log(1+x^{r_2}) \leq
   \log\left(1+\frac{x^{\beta_{22}-\gamma_{22}}}{1+\frac{x^{\beta_{21}-\gamma_{21}}}{1+x^{+\beta_{21}}}}\right)
   \label{eq:HK b}
\\&\log(1+x^{r_1}) +\log(1+x^{r_2}) \leq
  \log\left(1+\frac{x^{\beta_{22}-\gamma_{22}}+\frac{x^{\beta_{21}-\gamma_{21}}}{1+x^{-\beta_{21}}}}{1+\frac{x^{\beta_{21}-\gamma_{21}}}{1+x^{+\beta_{21}}}}\right)\nonumber\\
&+\log\left(1+\frac{\frac{x^{\beta_{11}-\gamma_{11}}}{1+x^{+\beta_{21}}}}{1+\frac{x^{\beta_{12}-\gamma_{12}}}{1+x^{+\beta_{12}}}}\right)\label{eq:HK c}
\\&\log(1+x^{r_1}) +\log(1+x^{r_2}) \leq
  \log\left(1+\frac{x^{\beta_{11}-\gamma_{11}}+\frac{x^{\beta_{12}-\gamma_{12}}}{1+x^{-\beta_{12}}}}{1+\frac{x^{\beta_{12}-\gamma_{12}}}{1+x^{+\beta_{12}}}}\right)\nonumber\\
&+\log\left(1+\frac{\frac{x^{\beta_{22}-\gamma_{22}}}{1+x^{+\beta_{12}}}}{1+\frac{x^{\beta_{21}-\gamma_{21}}}{1+x^{+\beta_{21}}}}\right)\label{eq:HK d}
\\&\log(1+x^{r_1}) +\log(1+x^{r_2}) \leq
  \log\left(1+\frac{\frac{x^{\beta_{11}-\gamma_{11}}}{1+x^{+\beta_{21}}}+\frac{x^{\beta_{12}-\gamma_{12}}}{1+x^{-\beta_{12}}}}{1+\frac{x^{\beta_{12}-\gamma_{12}}}{1+x^{+\beta_{12}}}}\right)\nonumber\\
&+\log\left(1+\frac{\frac{x^{\beta_{22}-\gamma_{22}}}{1+x^{+\beta_{12}}}+\frac{x^{\beta_{21}-\gamma_{21}}}{1+x^{-\beta_{21}}}}{1+\frac{x^{\beta_{21}-\gamma_{21}}}{1+x^{+\beta_{21}}}}\right)
\label{eq:HK e}
\\&2\log(1+x^{r_1}) +\log(1+x^{r_2}) \leq
    \log\left(1+\frac{x^{\beta_{11}-\gamma_{11}}+\frac{x^{\beta_{12}-\gamma_{12}}}{1+x^{-\beta_{12}}}}{1+\frac{x^{\beta_{12}-\gamma_{12}}}{1+x^{+\beta_{12}}}}\right)\nonumber
\\&+\log\left(1+\frac{\frac{x^{\beta_{11}-\gamma_{11}}}{1+x^{+\beta_{21}}}}{1+\frac{x^{\beta_{12}-\gamma_{12}}}{1+x^{+\beta_{12}}}}\right)
+\log\left(1+\frac{\frac{x^{\beta_{22}-\gamma_{22}}}{1+x^{+\beta_{12}}}+\frac{x^{\beta_{21}-\gamma_{21}}}{1+x^{-\beta_{21}}}}{1+\frac{x^{\beta_{21}-\gamma_{21}}}{1+x^{+\beta_{21}}}}\right)
\label{eq:HK f}
\\&\log(1+x^{r_1}) +2\log(1+x^{r_2}) \leq
    \log\left(1+\frac{x^{\beta_{22}-\gamma_{22}}+\frac{x^{\beta_{21}-\gamma_{21}}}{1+x^{-\beta_{21}}}}{1+\frac{x^{\beta_{21}-\gamma_{21}}}{1+x^{+\beta_{21}}}}\right)\nonumber
\\&+
\log\left(1+\frac{\frac{x^{\beta_{22}-\gamma_{22}}}{1+x^{+\beta_{12}}}}{1+\frac{x^{\beta_{21}-\gamma_{21}}}{1+x^{+\beta_{21}}}}\right)
+\log\left(1+\frac{\frac{x^{\beta_{11}-\gamma_{11}}}{1+x^{+\beta_{21}}}+\frac{x^{\beta_{12}-\gamma_{12}}}{1+x^{-\beta_{12}}}}{1+\frac{x^{\beta_{12}-\gamma_{12}}}{1+x^{+\beta_{12}}}}\right)
\bigg\}.\label{eq:HK g}
\end{align}
\label{eq:HK}
\end{subequations}

\subsection{Diversity}
The probability of outage $\mathbb{P}_{\rm out}(r_1,r_2)$ is
defined as the probability that the fading realization
$(\gamma_{11},\gamma_{12},\gamma_{21},\gamma_{22})$
is such that the rate pair $(r_1,r_2)$ cannot be decoded.
By using the outer bound region $\mathcal{R}_{\rm ETW}$ in~\reff{eq:outerbound within one bit}
and the inner bound region $\mathcal{R}_{\rm HK}$ in~\reff{eq:HK} we can bound the
outage probability as
\begin{align*}
  1-\mathbb{P}[(\gamma_{11},\gamma_{12},\gamma_{21},\gamma_{22})\in\mathcal{R}_{\rm ETW}] \leq
  \mathbb{P}_{\rm out}(r_1,r_2)
  \leq 1-\mathbb{P}[(\gamma_{11},\gamma_{12},\gamma_{21},\gamma_{22})\in\mathcal{R}_{\rm HK}].
\end{align*}
The diversity, or the high-SNR exponent of the outage probability, is defined as
\begin{align*}
d(r_1,r_2)
  &=    \lim_{x\to+\infty}\frac{-\log(\mathbb{P}_{\rm out}(r_1,r_2))}{\log(x)},
\end{align*}
and it is bounded by
\begin{align}
d_{\rm HK}(r_1,r_2)
\leq d(r_1,r_2)
\leq d_{\rm ETW}(r_1,r_2),
\label{eq:divivdiv}
\end{align}
where $d_{\rm ETW}(r_1,r_2)$ and $d_{\rm HK}(r_1,r_2)$ are defined similarly to $d(r_1,r_2)$.

The rest of the paper is devoted to the evaluation of $d_{\rm ETW}(r_1,r_2)$ and $d_{\rm HK}(r_1,r_2)$.

\section{Diversity upper bound}
\label{sec:upperbound}
By using the Laplace's integration method as in~\cite{zheng-tse:dmt}
we obtain
\begin{align}
d_{\rm ETW}(r_1,r_2)
  &= \min_{\gamma\in(\widetilde{\mathcal{R}}_{\rm ETW})^c}
  \{\gamma_{11}+\gamma_{12}+\gamma_{21}+\gamma_{22}\}
\label{eq:d etw}
\end{align}
where $\widetilde{\mathcal{R}}_{\rm ETW}$ is the large-$x$ approximation of
$\mathcal{R}_{\rm ETW}$ in~\reff{eq:outerbound within one bit} and is given by
\begin{subequations}
\begin{align}
&\widetilde{\mathcal{R}}_{\rm ETW} =
\bigg\{(\gamma_{11},\gamma_{12},\gamma_{21},\gamma_{22})\in\RR^4_+:
\quad X_{ij} \defeq [\beta_{ij}-\gamma_{ij}]^+, \nonumber
%
\\&r_1 \leq X_{11}
   \label{eq:a lim}
\\&r_2 \leq X_{22}
   \label{eq:b lim}
\\&r_s\defeq r_1 + r_2 \leq  [X_{11}-X_{21}]^+ +\max\{X_{21},X_{22}\} \nonumber
\\& \qquad=   \max\{X_{11},X_{21}\} +\max\{X_{22},X_{21}\}-X_{21}
   \label{eq:c lim}
\\&r_s\defeq r_1 + r_2 \leq  [X_{22}-X_{12}]^+ +\max\{X_{12},X_{11}\} \nonumber
\\& \qquad=   \max\{X_{11},X_{12}\} +\max\{X_{22},X_{12}\}-X_{12}
   \label{eq:d lim}
\\&r_s\defeq r_1 + r_2 \leq
    \max\{X_{12},X_{11} - X_{21}\}
   +\max\{X_{21},X_{22} - X_{12}\} \nonumber
\\& \qquad=   \max\{X_{11},X_{21}+X_{12}\} +\max\{X_{22},X_{21}+X_{12}\}-(X_{21}+X_{12}) \label{eq:e lim}
\\&r_f\defeq 2r_1 + r_2\leq
    [X_{11}-X_{21}]^+ +\max\{X_{11},X_{12}\}
    +\max\{X_{21},X_{22}-X_{12}\} \nonumber
\\& \qquad=   \max\{X_{11},X_{21}\}+\max\{X_{11},X_{12}\}
    +\max\{X_{22},X_{21}+X_{12}\}-(X_{21}-X_{12})
    \label{eq:f lim}
\\&r_g\defeq  r_1 + 2r_2\leq
    [X_{22}-X_{12}]^+ + \max\{X_{22},X_{21}\}
   +\max\{X_{12},X_{11}-X_{21}\} \nonumber
\\& \qquad=   \max\{X_{22},X_{21}\}+\max\{X_{22},X_{12}\}
    +\max\{X_{11},X_{21}+X_{12}\}-(X_{21}-X_{12}) \bigg\}
   \label{eq:g lim}
\end{align}
\label{eq:ETW lim}
\end{subequations}
where $[x]^+\defeq \max\{0,x\}$.

The optimization problem in~\reff{eq:d etw} can be solved as follows:
since the complement of $\widetilde{\mathcal{R}}_{\rm ETW}$ is the union of
the complement of the conditions~\reff{eq:a lim} through~\reff{eq:g
lim}, by applying the union bound as in~\cite{tse:dmt-mac} it can be
shown that the diversity in~\reff{eq:d etw} evaluates to
\begin{align*}
d_{\rm ETW}(r_1,r_2)
  &= \min_{\ell=a...g}\{d_{(\ref{eq:ETW lim}\ell)}\},
\\
d_{(\ref{eq:ETW lim}\ell)}
  &\defeq \beta_{11}+\beta_{12}+\beta_{21}+\beta_{22}
 -\max_{X\text{'s do NOT satisfy equation ($\ref{eq:ETW lim}\ell$)}}
\{X_{11}+X_{12}+X_{21}+X_{22}\}.
\end{align*}
We have:
\begin{itemize}
\item
The diversity $d_{\reff{eq:a lim}}$ (corresponding to the constraint~\reff{eq:a lim}) is:
\begin{align*}
d_{\reff{eq:a lim}}
  =&\beta_{11}-\max\{X_{11}\}
\\ &\textrm{subj. to} \,\, 0\leq X_{11} \leq \beta_{11}, \quad X_{11}\leq  r_1,
\\=&\beta_{11}- \min\{\beta_{11},r_1\}=-\min\{0,r_1-\beta_{11}\}=\max\{0,\beta_{11}-r_1\}
\\=& [\beta_{11}- r_1]^+.
\end{align*}

\item
Similarly to $d_{\reff{eq:a lim}}$, the diversity $d_{\reff{eq:b lim}}$ (corresponding to the constraint~\reff{eq:b lim}) is:
\begin{align*}
d_{\reff{eq:b lim}} = [\beta_{22}-r_2]^+.
\end{align*}

\item
The diversity $d_{\reff{eq:c lim}}$ (corresponding to the constraint~\reff{eq:c lim}) is:
\begin{align*}
d_{\reff{eq:c lim}}
  =&\beta_{11}+\beta_{21}+\beta_{22}-\max\{X_{11}+X_{21}+X_{22}\}
\\ &\textrm{subj. to}
 \,\, 0\leq X_{11} \leq \beta_{11},
 \,\, 0\leq X_{21} \leq \beta_{21},
 \,\, 0\leq X_{22} \leq \beta_{22},
\\ &\textrm{and to} \,\, \max\{X_{11},X_{21}\} + \max\{X_{22},X_{21}\} - X_{21} \leq r_s\defeq r_1+r_2.
\end{align*}
We start by re-writing the last constraint as follows:
\[
 \max\{X,Y\} + \max\{Z,Y\}  \leq r_s+ Y
 \Longleftrightarrow
 \left\{\begin{array}{l}
 X+Z \leq r_s+ Y\\
 Y+Z \leq r_s+ Y\\
 X+Y \leq r_s+ Y\\
 Y+Y \leq r_s+ Y\\
 \end{array}\right. \Longleftrightarrow
 \left\{\begin{array}{l}
 X+Z \leq r_s+ Y\\
 \max\{X,Y,Z\} \leq r_s\\
 \end{array}\right.,
\]
which implies
\begin{align*}
   &\max\{X_{21}+(X_{11}+X_{22})\}
\\ &\textrm{subj. to}
\,\, 0\leq X_{21} \leq \min\{r_s,\beta_{21}\},
\,\, 0\leq X_{11} \leq \min\{r_s,\beta_{11}\},
\,\, 0\leq X_{22} \leq \min\{r_s,\beta_{22}\},
\\ &\textrm{and to}
\,\, X_{11}+X_{22} \leq r_s+X_{21},
\\=&\max\Big\{X_{21}+ \min\Big\{\min\{r_s,\beta_{11}\}+\min\{r_s,\beta_{22}\}, r_s+X_{21} \Big\} \Big\}
\\ &\textrm{subj. to}
\,\, 0\leq X_{21} \leq \min\{r_s,\beta_{21}\},
\\=&\min\Big\{\min\{r_s,\beta_{11}\}+\min\{r_s,\beta_{22}\}+\min\{r_s,\beta_{21}\},
   \,\, r_s+2\min\{r_s,\beta_{21}\} \Big\}.
\end{align*}
Hence we obtain:
\begin{align*}
d_{\reff{eq:c lim}}
  &=\beta_{11}+\beta_{21}+\beta_{22}
\\&-\min\Big\{\min\{r_s,\beta_{11}\}+\min\{r_s,\beta_{22}\}+\min\{r_s,\beta_{21}\},
   r_s+2\min\{r_s,\beta_{21}\} \Big\}
\\&= \max\{
 [\beta_{11}-r_s]^+
+[\beta_{22}-r_s]^+
+[\beta_{21}-r_s]^+,
\beta_{11}+\beta_{22}-2r_s
+|\beta_{21}-r_s|
\}.
\end{align*}

Remark: In the symmetric case, with $\beta_{11}=\beta_{22}=1$,
$\beta_{12}=\beta_{21}=\alpha$ and $r_1=r_2=r$ (i.e., $r_s=2r$),
$d_{\reff{eq:c lim}}$ reduces to
\begin{align*}
d_{\reff{eq:c lim},sym}
&= 2\,\max\Big\{[1-2r]^++[\frac{\alpha}{2}-r]^+,(1-2r)+\left|\frac{\alpha}{2}-r\right|\Big\} \\
&= 2\, \left([A]^++[B]^+ +\Big[[B]^- -[A]^- \Big]^+\right)|_{A=1-2r, B=\frac{\alpha}{2}-r},
\end{align*}
where
\[
x = [x]^+ - [x]^-:
\quad [x]^+\defeq \max\{0,x\}\geq 0,
\quad [x]^-\defeq-\min\{0,x\}\geq 0.
\quad \forall x\in\RR.
\]
The corresponding bound in~\cite{leveque:izs2008} is
\begin{align*}
d_b
=&2[1-r-\min(r,\frac{\alpha}{2})]^++[\alpha-2r]^+\\
&= 2\, \left([A]^++[B]^+ +\Big[[A-B]^+ -[A]^+ +[B]^+\Big]^+\right)|_{A=1-2r, B=\frac{\alpha}{2}-r},
\end{align*}
which can be easily shown to be equivalent to $d_{\reff{eq:c lim},sym}$ since for
$A\geq B: \,\ [A-B]^+-[A]^++[B]^+=A-B-[A]^++[B]^+ = -[A]^- +[B]^-$, and
$A< B: \,\ [A-B]^+-[A]^++[B]^+=-[A]^++[B]^+\leq 0$.

\item
The diversity $d_{\reff{eq:d lim}}$ (corresponding to the constraint~\reff{eq:d lim}) is as
$d_{\reff{eq:c lim}}$ but with $\beta_{21}$ replaced by $\beta_{12}$, i.e., with the role of the users swapped.

Remark: In the symmetric case, $d_{\reff{eq:d lim},sym}=d_{\reff{eq:c lim},sym}$.

\item
The diversity $d_{\reff{eq:e lim}}$ (corresponding to the constraint~\reff{eq:e lim}) is as
$d_{\reff{eq:c lim}}$
but with $\beta_{21}+\beta_{12}$ instead of $\beta_{21}$.

Remark: In the symmetric case, $d_{\reff{eq:e lim},sym}$ coincides with $d_c$ in~\cite{leveque:izs2008}.



\item
The diversity $d_{\reff{eq:f lim}}$ (corresponding to the constraint~\reff{eq:f lim}) is:
\begin{align*}
d_{\reff{eq:f lim}}
  =&\beta_{11}+\beta_{12}+\beta_{21}+\beta_{22}-\max\{X+Y+Z+W\}
\\ &\textrm{subj. to}
 \,\, 0\leq X \leq \beta_{11},
 \,\, 0\leq W \leq \beta_{12},
 \,\, 0\leq Y \leq \beta_{21},
 \,\, 0\leq Z \leq \beta_{22},
\\ &\textrm{and to} \,\,
  \max\{X,Y\} + \max\{X,W\} + \max\{Z,Y+W\}-(Y+W) \leq r_f\defeq 2r_1+r_2.
\end{align*}
The last constraint can be rewritten as
\begin{align*}
   \left\{\begin{array}{l}
   (X-Y)+(X-W)\leq r_f - \max\{Z,Y+W\}\\
    X - Y     \leq r_f - \max\{Z,Y+W\} \\
    X - W     \leq r_f - \max\{Z,Y+W\} \\
        0     \leq r_f - \max\{Z,Y+W\} \\
   \end{array}\right.
\Longleftrightarrow
 \left\{\begin{array}{l}
   2X \leq r_f- [Z-(Y+W)]^+ \\
   \max\{Z,Y+W\} \leq r_f   \\
  \end{array}\right. ,
\end{align*}
that together with $ X \leq \beta_{11}$ gives
\[
\max\{X\} = \min\left\{\beta_{11}, \frac{r_f - [Z-(Y+W)]^+}{2}\right\}.
\]
Next, the function
\[
\max\{X\}+(Y+W)
= \min\left\{\beta_{11}, \frac{r_f - [Z-(Y+W)]^+}{2}\right\}+(Y+W),
\]
is increasing in $(Y+W)$, and since the constraints
\[
W \leq \beta_{12},\,\, Y \leq \beta_{21},\,\, Z \leq \beta_{22},\,\,\max\{Z,Y+W\} \leq r_f,
\]
are equivalent to
\[
Y+W \leq \min\{\beta_{21}+\beta_{12}, r_f\},\quad Z   \leq \min\{\beta_{22},r_f\},
\]
we have
\[
\max\{X+Y+W\}
=
\min\left\{\beta_{11}, \frac{r_f - [Z-\min\{r_f,\beta_{21}+\beta_{12}\}]^+}{2}\right\}+\min\{r_f,\beta_{21}+\beta_{12}\}.
\]
Finally, the function
\[
Z+\max\{X+Y+W\}
=
\min\left\{\beta_{11}, \frac{r_f - [Z-\min\{r_f,\beta_{21}+\beta_{12}\}]^+}{2}\right\}+\min\{r_f,\beta_{21}+\beta_{12}\}+Z,
\]
is also increasing in $Z$, hence, subject to $Z\leq\min\{r_f,\beta_{22}\}$,  we finally have
\begin{align*}
d_{\reff{eq:f lim}}
  &=-\max\{X+Y+W+Z\}+(\beta_{11}+\beta_{12}+\beta_{21}+\beta_{22})
\\&=
-\min\left\{\beta_{11}, \frac{r_f - [\min\{r_f,\beta_{22}\}-\min\{r_f,\beta_{21}+\beta_{12}\}]^+}{2}\right\}
-\min\{r_f,\beta_{21}+\beta_{12}\}
-\min\{r_f,\beta_{22}\}
\\&+(\beta_{11}+\beta_{12}+\beta_{21}+\beta_{22})
\\&=
\left[\beta_{11}-\frac{r_f - [\beta_{22}-(\beta_{21}+\beta_{12})
  +a-b]^+}{2}\right]^+ +a+b,
\quad
a\defeq [(\beta_{21}+\beta_{12})-r_f]^+,
b\defeq [\beta_{22}-r_f]^+.
\end{align*}

Remark: In the symmetric case,
$d_{\reff{eq:f lim}}$ reduces to
\begin{align*}
d_{\reff{eq:f lim},sym}= &\frac{1+b+[[a]^--[b]^-]^+}{2} +[a]^+ +[b]^+,
\quad a = 2\alpha-3r, b = 1-3r
\end{align*}
which is not equivalent to $d_d$ in~\cite{leveque:izs2008}.
In fact, it turns out that $d_d$ in~\cite{leveque:izs2008} is not correct.
Consider the following numerical example: let
$r_1=r_2=0.4$, $\beta_{11}=\beta_{22}=1$,
$\beta_{12}=\beta_{21}=0.5$, which corresponds to
$r_f=1.2, \alpha=0.5$  in~\cite{leveque:izs2008}.
The optimization problem for $d_d$ is
\begin{align*}
d_d&=\min\{\gamma_{11}+\gamma_{12}+\gamma_{21}+\gamma_{22}\}\\
\textrm{subj.to}&~[[1-\gamma_{11}]^+-[\alpha-\gamma_{12}]^+]^++\max([1-\gamma_{11}]^+,[\alpha-\gamma_{21}]^+)\\
&+\max([\alpha-\gamma_{12}]^+,[1-\gamma_{22}]^+-[\alpha-\gamma_{21}]^+)\leq r_f
\end{align*}
It can be easily verified that $\gamma_{11}=0.4$,
$\gamma_{12}=\gamma_{21}=\gamma_{22}=0$ is a feasible solution
that gives
$(\gamma_{11}+\gamma_{12}+\gamma_{21},\gamma_{22})=0.4$.
However, according
\begin{align*}
d_d=\max\{(1-\frac{3r}{2})^++(1-3r)^++(2\alpha-3r)^+,\min\{[3-3r-\min(3r,2\alpha)]^+,\max(1,2-3r-\min(3r,2\alpha))\}
\end{align*}
we have $d_d=0.8$.

\item
Finally $d_{\reff{eq:g lim}}$ (corresponding to the constraint~\reff{eq:g lim}) is as
$d_{\reff{eq:f lim}}$ but with $\beta_{22}$ instead of $\beta_{11}$
and $r_g$ instead of $r_f$, i.e., the role of the users is swapped.

\end{itemize}

\section{Diversity lower bound}
\label{sec:lowerbound}
The evaluation of the diversity lower bound $d_{\rm HK}$ in~\reff{eq:divivdiv}
can be carried out similarly to the evaluation of the diversity upper bound
$d_{\rm ETW}$ in the previous section.
We will consider both the case of no rate splitting and
the particular choice of the power split among common and private messages
inspired by~\cite{etkin_tse_hua:withinonebit:subIt06} which led to~\reff{eq:HK}.

\subsection{Diversity lower bound without rate splitting}
Without rate splitting in the HK region, a user either sends
all private information or all common information.  These two modes
of operation correspond to either treating the interference as noise
at the receiver, or performing joint decoding as in a MAC channel.

Consider first the case where the interference is treated as noise.
User~1 can be successfully decoded at receiver~1 by treating user~2 as noise if
\[
\{(\gamma_{11},\gamma_{12})\in\RR^2:\,\,
\log(1+x^{r_1})
\leq \log\left(1+\frac{x^{\beta_{11}-\gamma_{11}}}{1+x^{\beta_{12}-\gamma_{12}}}\right)\}.
\]
By following the same approach used in the derivation
of the diversity upper bound, we have that the exponent of the probability that
user~1 cannot be decoded successfully at receiver~1 by treating user~2 as noise
is given by:
\begin{align*}
d_{\rm  NI1}
   &=\beta_{11}+\beta_{12}-\max\{X_{11}+X_{12}\}
\\ &\textrm{subj. to}
 \,\, 0\leq X_{11} \leq \beta_{11},
 \,\, 0\leq X_{12} \leq \beta_{12},
 \,\,  [X_{11}-X_{12}]^+ \leq r_1
\\=&\beta_{11}+\beta_{12}-\max\{X_{11}+X_{12}\}
\\ &\textrm{subj. to}
 \,\, 0\leq X_{11} \leq \beta_{11},
 \,\, 0\leq X_{12} \leq \beta_{12},
 \,\, X_{11}\leq X_{12}+r_1
\\=&\beta_{11}+\beta_{12}-\max\{\min\{\beta_{11},X_{12}+r_1\}+X_{12}\}
\\ &\textrm{subj. to}
 \,\, 0\leq X_{12} \leq \beta_{12},
\\=&\beta_{11}+\beta_{12}-\max\{\min\{\beta_{11},\beta_{12}+r_1\}+\beta_{12}\}
\\&= [\beta_{11}-r_1-\beta_{12} ]^+,
\end{align*}
Similarly, the exponent of the probability that
user~2 cannot be successfully decoded at receiver~2 by treating user~1 as noise is
\[
d_{\rm NI2}= [\beta_{22}-r_2-\beta_{21} ]^+.
\]

Consider now the case where the users are jointly decoded.
User~1 and user~2 can be successfully jointly decoded at receiver~2 as
in a MAC if
\begin{align*}
\{
&(\gamma_{21},\gamma_{22})\in\RR^4:\\
&\log(1+x^{r_1})
   \leq \log(1+x^{\beta_{21}-\gamma_{21}})
\\
&\log(1+x^{r_2})
   \leq \log(1+x^{\beta_{22}-\gamma_{22}})
\\
&\log(1+x^{r_1})+\log(1+x^{r_2})
   \leq \log(1+x^{\beta_{21}-\gamma_{21}}+x^{\beta_{22}-\gamma_{22}})
\}.
\end{align*}
The exponent of the probability that both users cannot be jointly decoded is
given by
\[
d_{\rm MAC2} = \min\{
  [\beta_{21}-r_1]^+,
  [\beta_{22}-r_2]^+,
  [\beta_{22}-r_s]^+ + [\beta_{21}-r_s]^+
\}, \quad
r_s \defeq r_1+r_2,
\]
where the last argument of the minimum in $d_{\rm MAC2}$
can be derived as follows:
\begin{align*}
   &\beta_{22}+\beta_{21}-\max\{X_{21}+X_{22}\}
\\ &\textrm{subj. to}
 \,\, 0\leq X_{22} \leq \beta_{22},
 \,\, 0\leq X_{21} \leq \beta_{21},
 \,\, \max\{X_{22,}X_{21}\}\leq r_s,
\\=&\beta_{22}+\beta_{21}-\min\{\beta_{21},r_s\}-\min\{\beta_{22},r_s\}
\\=&[\beta_{22}-r_s]^+ + [\beta_{21}-r_s]^+.
\end{align*}
Similarly, the exponent of the probability that
user~1 and user~2 cannot be successfully jointly decoded at receiver~1 is
\[
d_{\rm MAC1} = \min\{
  [\beta_{11}-r_1]^+,
  [\beta_{12}-r_2]^+,
  [\beta_{12}-r_s]^+ + [\beta_{11}-r_s]^+
\}, \quad
r_s \defeq r_1+r_2.
\]

Hence, without rate splitting, we have
\begin{align}
d_{\rm HK-wors} = \max\{d_{00},d_{01},d_{10},d_{11}\},
\label{eq: d HK, wrs}
\end{align}
where ``wors'' stands for 	``without rate splitting'' and where
\begin{itemize}

\item
$d_{11}$ is the diversity when both sources send
only private information
(which is sum-rate optimal for very weak interference unfaded
GIFCs~\cite{shang:ifcout:it09}) given by
\begin{align*}
d_{11} = \min\{d_{\rm NI1},d_{\rm NI2}\}.
\end{align*}

\item
$d_{10}$ (and similarly for $d_{01}$ but the role of he users swapped)
is the diversity when user~1 sends only private information
and the user~2 sends only common information
(which is sum-rate optimal for mixed interference unfaded
GIFCs~\cite{tuninettiwengisit08}) given by
\begin{align*}
d_{10} = \min\{d_{\rm NI1},d_{\rm MAC2}\}.
\end{align*}

\item
$d_{00}$ is the diversity when both sources send common information
(which is optimal for strong interference unfaded
GIFCs~\cite{chen-kramer:isit2008-out}) given by
\begin{align*}
d_{00} = \min\{d_{\rm MAC1},d_{\rm MAC2}\}.
\end{align*}
\end{itemize}

Since $\max\{\min\{a,b_1\},\min\{a,b_2\}\}=\min\{a,\max\{b_1,b_2\}\}$, we further
rewrite $d_{\rm HK-wors}$ in~\reff{eq: d HK, wrs} as
\begin{align*}
d_{\rm HK-wors}
  &= \max\big\{
\min\{d_{\rm NI1} ,d_{\rm NI2}\},
\min\{d_{\rm NI1} ,d_{\rm MAC2}\},
\min\{d_{\rm MAC1},d_{\rm NI2}\},
\min\{d_{\rm MAC1},d_{\rm MAC2}\}
\big\}
\\&= \max\big\{
\min\{d_{\rm NI1} ,\max\{d_{\rm NI2},d_{\rm MAC2}\} \},
\min\{d_{\rm MAC1},\max\{d_{\rm NI2},d_{\rm MAC2}\} \}
\big\}
\\&=
\min\big\{\max\{d_{\rm NI1},d_{\rm MAC1}\},\, \max\{d_{\rm NI2},d_{\rm MAC2}\} \big\}
\\&=
\min\big\{d_{\rm wrp\,1},d_{\rm worp\,2} \big\},
\quad d_{{\rm worp}\,u}\defeq\max\{d_{{\rm NI}\,u},d_{{\rm MAC}\,u}\},
\end{align*}
that is, the diversity $d_{\rm HK-wors}$ has the following intuitive explanation:
each user $u\in\{1,2\}$ chooses the best strategy between treating the interference
as noise ($d_{{\rm NI}\,u}$) and joint decoding ($d_{{\rm MAC}\,u}$),
which gives diversity $d_{{\rm wrp}\,u}$,
and the overall diversity is dominated by the the worst user.

\subsection{Diversity lower bound with rate splitting}
In the general case of rate splitting, with the power split
indicated in Section~\ref{sec:chmodel}, we have:
\begin{align}
d_{\rm HK}(r_1,r_2)
  &= \min_{\gamma\in(\widetilde{\mathcal{R}}_{\rm HK})^c}
  \{\gamma_{11}+\gamma_{12}+\gamma_{21}+\gamma_{22}\}
\label{eq:d hk rate splitting}
\end{align}
where $\widetilde{\mathcal{R}}_{\rm HK}$ is the large-$x$ approximation of
$\mathcal{R}_{\rm HK}$ in~\reff{eq:HK} and is given by:
\begin{subequations}
\begin{align}
\widetilde{\mathcal{R}}_{\rm HK} =
\bigg\{
&(\gamma_{11},\gamma_{12},\gamma_{21},\gamma_{22})\in\RR^4_+:
\quad X_{ij} \defeq \beta_{ij}-\gamma_{ij}, \nonumber
\\&r_1\leq [X_{11}]^+  \label{eq:HK sim a lim}
\\&r_2\leq [X_{22}]^+  \label{eq:HK sim b lim}
\\&r_s\defeq r_1+r_2 \leq
  [\max\{X_{22}, X_{21}\}]^+ +[X_{11}-\beta_{21}]^+
\label{eq:HK sim c lim}
\\&r_s\defeq r_1+r_2 \leq
  [\max\{X_{11}, X_{12}\}]^+ +[X_{22}-\beta_{12} ]^+
\label{eq:HK sim d lim}
\\&r_s\defeq r_1+r_2 \leq
  [\max\{X_{11}-\beta_{21}, X_{12}\}]^+ +[\max\{X_{22}-\beta_{12}, X_{21}\}]^+
\label{eq:HK sim e lim}
\\&r_f\defeq 2r_1+r_2 \leq
  [\max\{X_{11}, X_{12}\}]^+ +[X_{11}-\beta_{21}]^+\nonumber\\
&\quad+[\max\{X_{22}-\beta_{12},X_{21}\}]^+
\label{eq:HK sim f lim}
\\&r_g\defeq r_1+2r_2 \leq
  [\max\{X_{22},X_{21}\}]^+ +[X_{22}-\beta_{12}]^+\nonumber\\
&\quad+[\max\{X_{11}-\beta_{21},X_{12}\}]^+ \bigg\}.
\label{eq:HK sim g lim}
\end{align}
\label{eq:HK sim lim}
\end{subequations}

We finally have that the diversity is given by:
\begin{align*}
d_{\rm HK}
 &= \min_{\ell= a...g}\{d_{(\ref{eq:HK sim lim}\ell)}\},
\\
d_{(\ref{eq:HK sim lim}\ell)}
  &\defeq \beta_{11}+\beta_{12}+\beta_{21}+\beta_{22}
 -\max_{X\text{'s do NOT satisfy equation (\ref{eq:HK sim lim}\,$\ell$)}}
\{X_{11}+X_{12}+X_{21}+X_{22}\}.
\end{align*}
where
\begin{itemize}

\item
The diversity $d_{\reff{eq:HK sim a lim}}$  and $d_{\reff{eq:HK sim b lim}}$
(corresponding to the constraint~\reff{eq:HK sim a lim} and~\reff{eq:HK sim b lim},
respectively) are:
\[
d_{\reff{eq:HK sim a lim}}
= d_{\reff{eq:a lim}}
= [\beta_{11}-r_1]^+,
\]
and
\[
d_{\reff{eq:HK sim b lim}}
= d_{\reff{eq:b lim}}
=  [\beta_{22}-r_2]^+,
\]
as for the upper bound.

\item
The diversity $d_{\reff{eq:HK sim c lim}}$ (corresponding to the
constraint~\reff{eq:HK sim c lim}) is:
\begin{align*}
d_{\reff{eq:HK sim c lim}}
=&\beta_{22}+\beta_{21}+\beta_{11}
  -\max\{X_{22}+X_{21}+X_{11}\}
\\ &\textrm{subj. to}
   \,\, X_{22}\leq \beta_{22},
   \,\, X_{21}\leq \beta_{21},
   \,\, X_{11}\leq \beta_{11},
\\&\textrm{and to}
   \,\, [\max\{X_{22},X_{21}\}]^+ +[X_{11}-\beta_{21}]^+ \leq r_s\defeq r_1+r_2.
\end{align*}
If rewrite the optimization domain as
\begin{align*}
  &X_{22}\leq \beta_{22}, \,\,X_{21}\leq \beta_{21}, \,\,
  \max\{X_{22},X_{21}\} \leq \alpha r_s,
\\&X_{11}\leq \beta_{11}, \,\,X_{11}-\beta_{21}     \leq (1-\alpha) r_s,
\\&  \alpha\in[0,1],
\end{align*}
we immediately obtain
\begin{align*}
d_{\reff{eq:HK sim c lim}}
  &= \beta_{22}+\beta_{21}+\beta_{11}-\max_{\alpha\in[0,1]}\Big\{
 \min\{\beta_{22},\alpha r_s\}
+\min\{\beta_{21},\alpha r_s\}
+\min\{\beta_{11},(1-\alpha) r_s+\beta_{21}\} \Big\}
\\&=  \min_{\alpha\in[0,1]}\{
  [\beta_{22}-\alpha r_s]^+
+ [\beta_{21}-\alpha r_s]^+
+ [\beta_{11}-\beta_{21}-(1-\alpha) r_s]^+ \},
\end{align*}
and the optimal $\alpha$ (see Appendix~\ref{app:alpha fc}) is
\[
\alpha_{\reff{eq:HK sim c lim}}\defeq \min\left\{1,\max\{\beta_{22},\beta_{21}\}/r_s\right\}.
\]


Remark: In general, it is difficult to compare the lower bound $d_{\reff{eq:HK sim c lim}}$
with the upper bound $d_{\reff{eq:c lim}}$.  By inspection, if $r_s\geq\beta_{21}\geq\max\{\beta_{11},\beta_{22}\}$
the two bounds meet. It is possible that this condition we can be relaxed.

\item
The diversity $d_{\reff{eq:HK sim d lim}}$ (corresponding to the
constraint~\reff{eq:HK sim d lim}) is as $d_{\reff{eq:HK sim c lim}}$ but with the role of the users reversed.

\item
The diversity $d_{\reff{eq:HK sim e lim}}$ (corresponding to the
constraint~\reff{eq:HK sim e lim}) is:
\begin{align*}
d_{\reff{eq:HK sim e lim}}
=&\beta_{22}+\beta_{21}+\beta_{11}+\beta_{12}
  -\max\{X_{22}+X_{21}+X_{11}+X_{12}\}
\\ &\textrm{subj. to}
   \,\, X_{22}\leq \beta_{22},
   \,\, X_{21}\leq \beta_{21},
   \,\, X_{11}\leq \beta_{11},
   \,\, X_{12}\leq \beta_{12},
\\&\textrm{and to}
   \,\, [\max\{X_{11}-\beta_{21},X_{12}\}]^+ +  [\max\{X_{22}-\beta_{12},X_{21}\}]^+\leq r_s.
\end{align*}
If rewrite the optimization domain as
\begin{align*}
   &    X_{22}\leq \beta_{22},
   \,\, X_{21}\leq \beta_{21},
   \,\,  \max\{X_{22}-\beta_{12},X_{21}\} \leq \alpha r_s,
\\&     X_{11}\leq \beta_{11},
   \,\, X_{12}\leq \beta_{12},
   \,\,  \max\{X_{11}-\beta_{21},X_{12}\} \leq  (1-\alpha) r_s,
\\&  \alpha\in[0,1],
\end{align*}
we immediately obtain
\begin{align*}
d_{\reff{eq:HK sim e lim}}
&=
\beta_{22}+\beta_{21}+\beta_{11}+\beta_{12}
\\& -\max_{\alpha\in[0,1]}\Big\{
  \min\{\beta_{22},\beta_{12}+\alpha r_s\}
 +\min\{\beta_{21},           \alpha r_s\}
 +\min\{\beta_{11},\beta_{21}+(1-\alpha) r_s\}
 +\min\{\beta_{12},           (1-\alpha) r_s\}
  \Big\}
\\&=\min_{\alpha\in[0,1]}\Big\{
  [\beta_{22}-\beta_{12}-\alpha r_s]^+
 +[\beta_{21}-           \alpha r_s]^+
 +[\beta_{11}-\beta_{21}-(1-\alpha) r_s]^+
 +[\beta_{12}-           (1-\alpha) r_s]^+
  \Big\},
\end{align*}
and the optimal $\alpha$ (see Appendix~\ref{app:alpha fe}) is
\[
\alpha_{\reff{eq:HK sim c lim}}
\defeq
 \min\left\{1,\frac{\max\{\beta_{22}-\beta_{21},\beta_{21}\}}{r_s}\right\}.
\]

\item
The diversity $d_{\reff{eq:HK sim f lim}}$ (corresponding to the
constraint~\reff{eq:HK sim f lim}) is:
\begin{align*}
d_{\reff{eq:HK sim f lim}}
=&\beta_{22}+\beta_{21}+\beta_{11}+\beta_{12}
  -\max\{X_{22}+X_{21}+X_{11}+X_{12}\}
\\ &\textrm{subj. to}
   \,\, X_{22}\leq \beta_{22},
   \,\, X_{21}\leq \beta_{21},
   \,\, X_{11}\leq \beta_{11},
   \,\, X_{12}\leq \beta_{12},
\\&\textrm{and to}
   \,\, [\max\{X_{11}, X_{12}\}]^+ +[X_{11}-\beta_{21}]^+
   +[\max\{X_{22}-\beta_{12},X_{21}\}]^+ \leq r_f
\end{align*}
If rewrite the optimization domain as
\begin{align*}
   &    X_{22}\leq \beta_{22},
   \,\, X_{21}\leq \beta_{21},
   \,\,  \max\{X_{22}-\beta_{12},X_{21}\} \leq \alpha r_f,
\\&     X_{11}\leq \beta_{11},
   \,\, X_{12}\leq \beta_{12},
   \,\,  \max\{X_{11}, X_{12}\} +[X_{11}-\beta_{21}]^+ \leq  (1-\alpha) r_f,
\\&  \alpha\in[0,1],
\end{align*}
and $\max\{X_{11}, X_{12}\} +[X_{11}-\beta_{21}]^+ \leq  (1-\alpha) r_f$ as
\begin{align*}
  &X_{11}+X_{11}     \leq \beta_{21}+(1-\alpha) r_f
\\&X_{11}+\beta_{21} \leq \beta_{21}+(1-\alpha) r_f
\\&X_{12}+X_{11}     \leq \beta_{21}+(1-\alpha) r_f
\\&X_{12}+\beta_{21} \leq \beta_{21}+(1-\alpha) r_f
\end{align*}
we obtain that the optimization domain is
\begin{align*}
   &    X_{22}\leq \min\{\beta_{22},\beta_{12}+\alpha r_f\},
   \,\, X_{21}\leq \min\{\beta_{21},\alpha r_f \},
\\&     X_{11}\leq \min\{\beta_{11},(1-\alpha) r_f,\frac{\beta_{21}+(1-\alpha) r_f}{2}\},
   \,\, X_{12}\leq \min\{\beta_{12},(1-\alpha) r_f\},
\\&     X_{11}+X_{12} \leq  \beta_{21}+(1-\alpha) r_f,
   \,\,\alpha\in[0,1],
\end{align*}
and we immediately obtain
\begin{align*}
d_{\reff{eq:HK sim f lim}}
&=
\beta_{22}+\beta_{21}+\beta_{11}+\beta_{12}
-\max_{\alpha\in[0,1]}\Big\{
  \min\{\beta_{22},\beta_{12}+\alpha r_f\}
 +\min\{\beta_{21},           \alpha r_f\}
\\&  +\min\Big\{\beta_{21}+(1-\alpha) r_f, \,\,\,\,
 \min\{\beta_{11},(1-\alpha) r_f,\frac{\beta_{21}+(1-\alpha) r_f}{2}\}+\min\{\beta_{12},(1-\alpha) r_f\}
 \Big\}
 \Big\},
\end{align*}
and the optimal $\alpha$ can be found in Appendix~\ref{app:HK gen}.

\item
The diversity $d_{\reff{eq:HK sim g lim}}$ (corresponding to the
constraint~\reff{eq:HK sim g lim}) is as $d_{\reff{eq:HK sim e lim}}$ but with the role of the users reversed.

\end{itemize}

Remark: In this section we derived the achievable diversity for the case of power split
such that the average power of the private message at the non-intended receiver is
below the noise floor.
In Appendix~\ref{app:HK gen}, we derive the achievable diversity for a generic power split.
The expression is quite complex and not very insightful.
By numerical optimization we found that the particular power split we chose in Section~\ref{sec:chmodel}
is optimal, or very close to optimal, for a very large set of channel parameters
$(\beta_{11}, \beta_{12}, \beta_{21}, \beta_{22})$.

\section{Numerical results}
\label{sec:num}

In this section we present numerical evaluations of the
diversity upper bound $d_{\rm ETW}$ in~\reff{eq:d etw} and the diversity
lower bound without rate splitting $d_{\rm HK-wors}$ in~\reff{eq: d HK, wrs}
and the diversity lower bound with rate splitting $d_{\rm HK}$ in~\reff{eq:d hk rate splitting}
for different values of
the channel parameters $(\beta_{11}, \beta_{12}, \beta_{21}, \beta_{22})$.

\subsection{Symmetric channels}

    \begin{figure}
        \begin{center}
        \includegraphics[width=12cm]{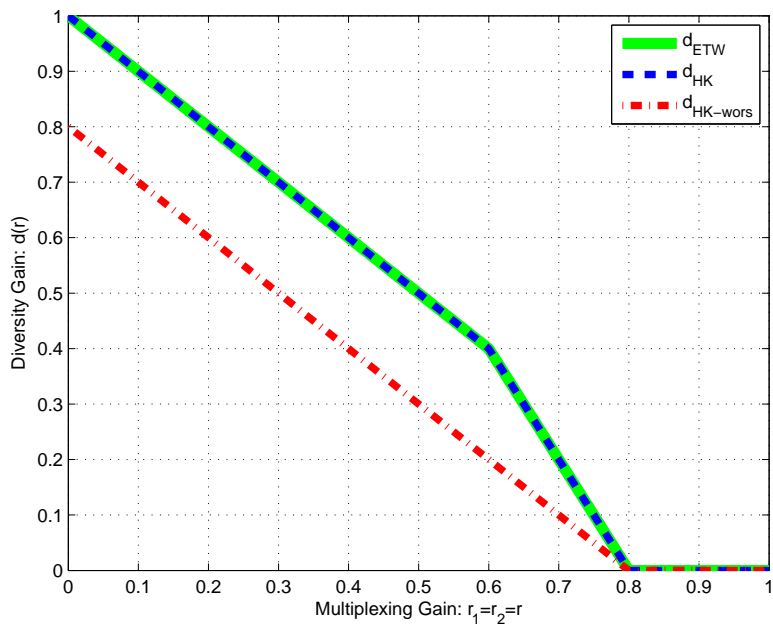}
        \caption{Symmetric channel in weak interference: $\beta_{11}=\beta_{22}=1, \beta_{12}=\beta_{21}=0.2$.}
        \label{fig:sumrate0.2}
        \end{center}
    \end{figure}

    \begin{figure}
        \begin{center}
        \includegraphics[width=12cm]{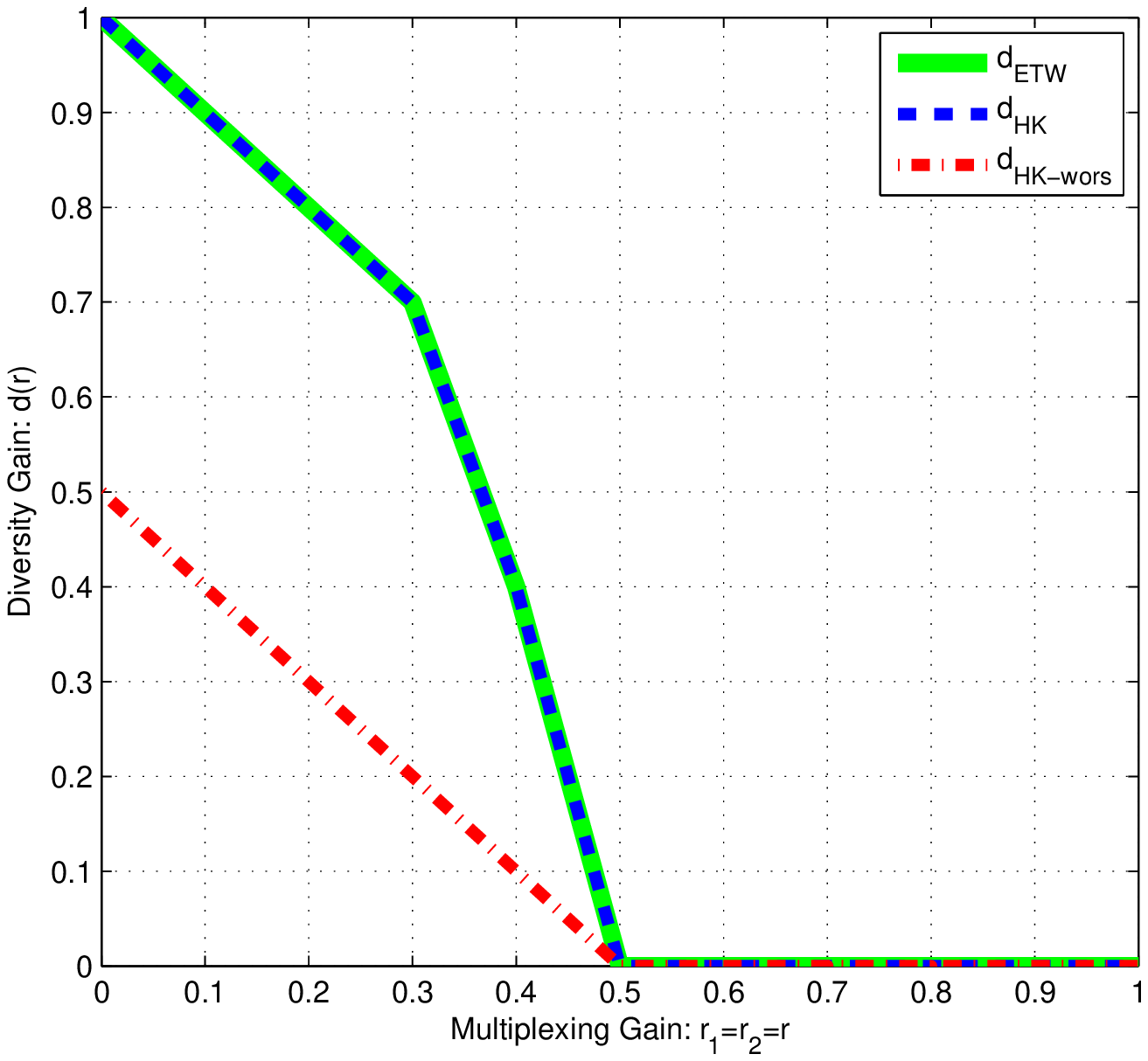}
        \caption{Symmetric channel in weak interference: diversity vs. $r_1=r_2=r$ for $\beta_{11}=\beta_{22}=1, \beta_{12}=\beta_{21}=0.5$.}
        \label{fig:sumrate0.5}
        \end{center}
    \end{figure}

    \begin{figure}
        \begin{center}
        \includegraphics[width=12cm]{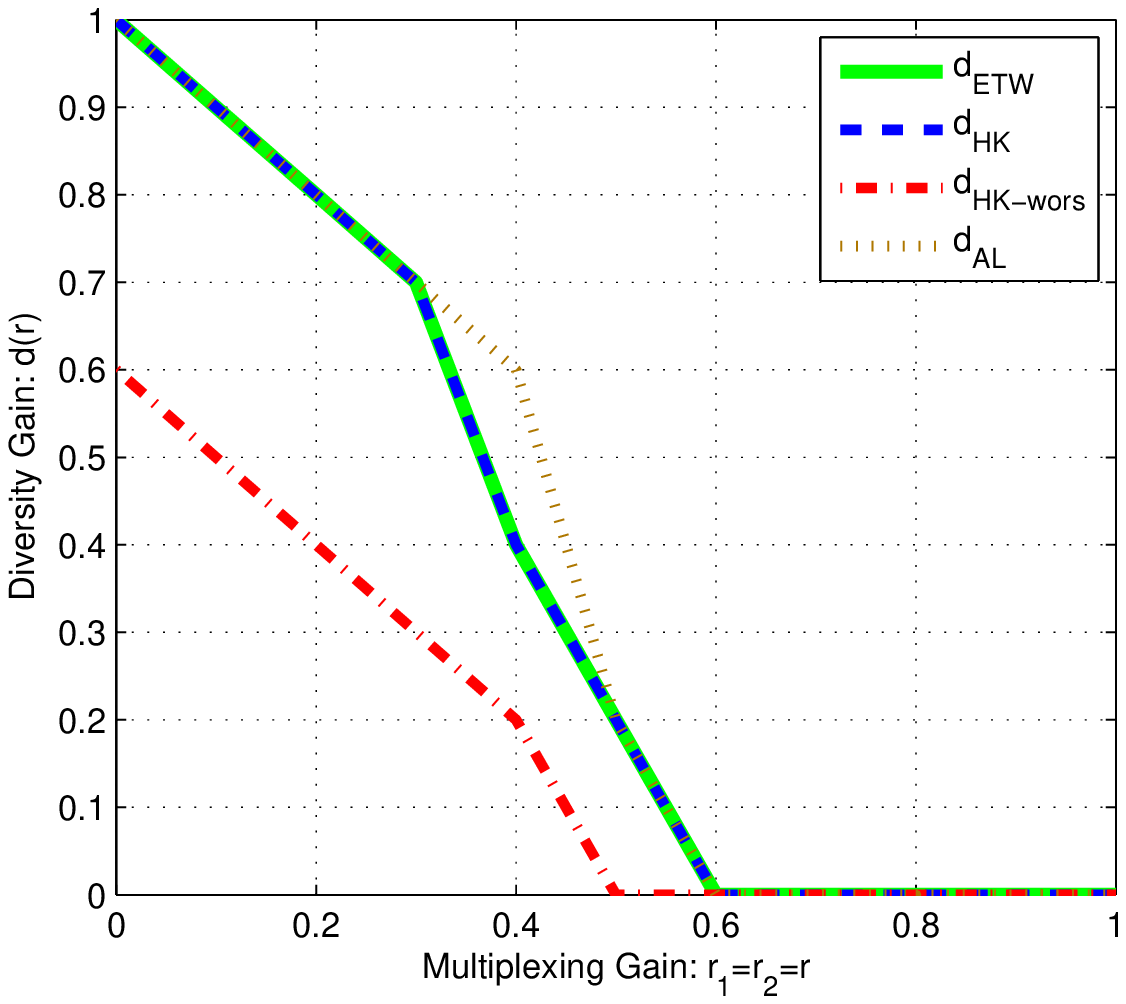}
        \caption{Symmetric channel in weak interference: diversity vs. $r_1=r_2=r$ for $\beta_{11}=\beta_{22}=1, \beta_{12}=\beta_{21}=0.6$.}
        \label{fig:sumrate0.6}
        \end{center}
    \end{figure}

    \begin{figure}
        \begin{center}
        \includegraphics[width=12cm]{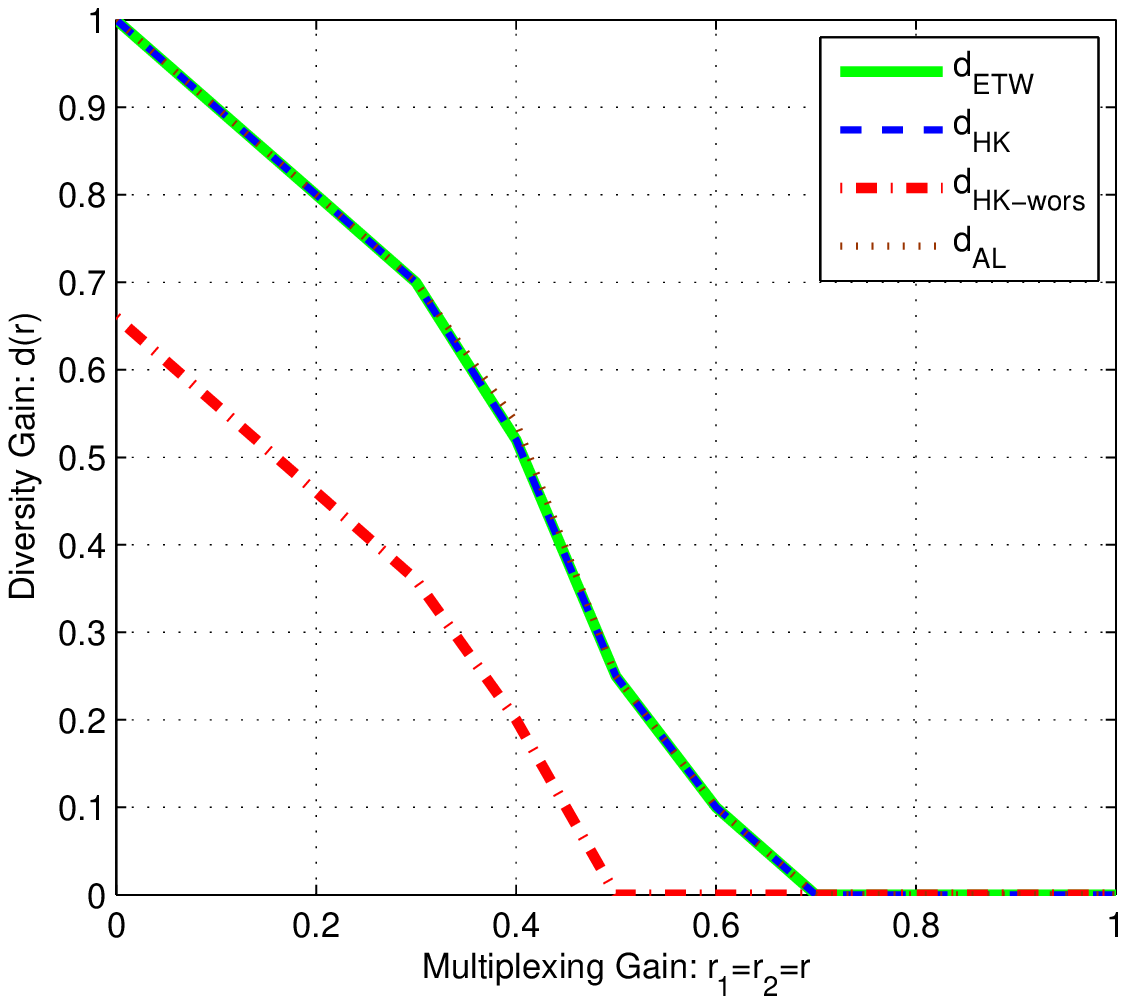}
        \caption{Symmetric channel in weak interference: diversity vs. $r_1=r_2=r$ for $\beta_{11}=\beta_{22}=1, \beta_{12}=\beta_{21}=2/3$.}
        \label{fig:sumrate0.66}
        \end{center}
    \end{figure}

    \begin{figure}
        \begin{center}
        \includegraphics[width=12cm]{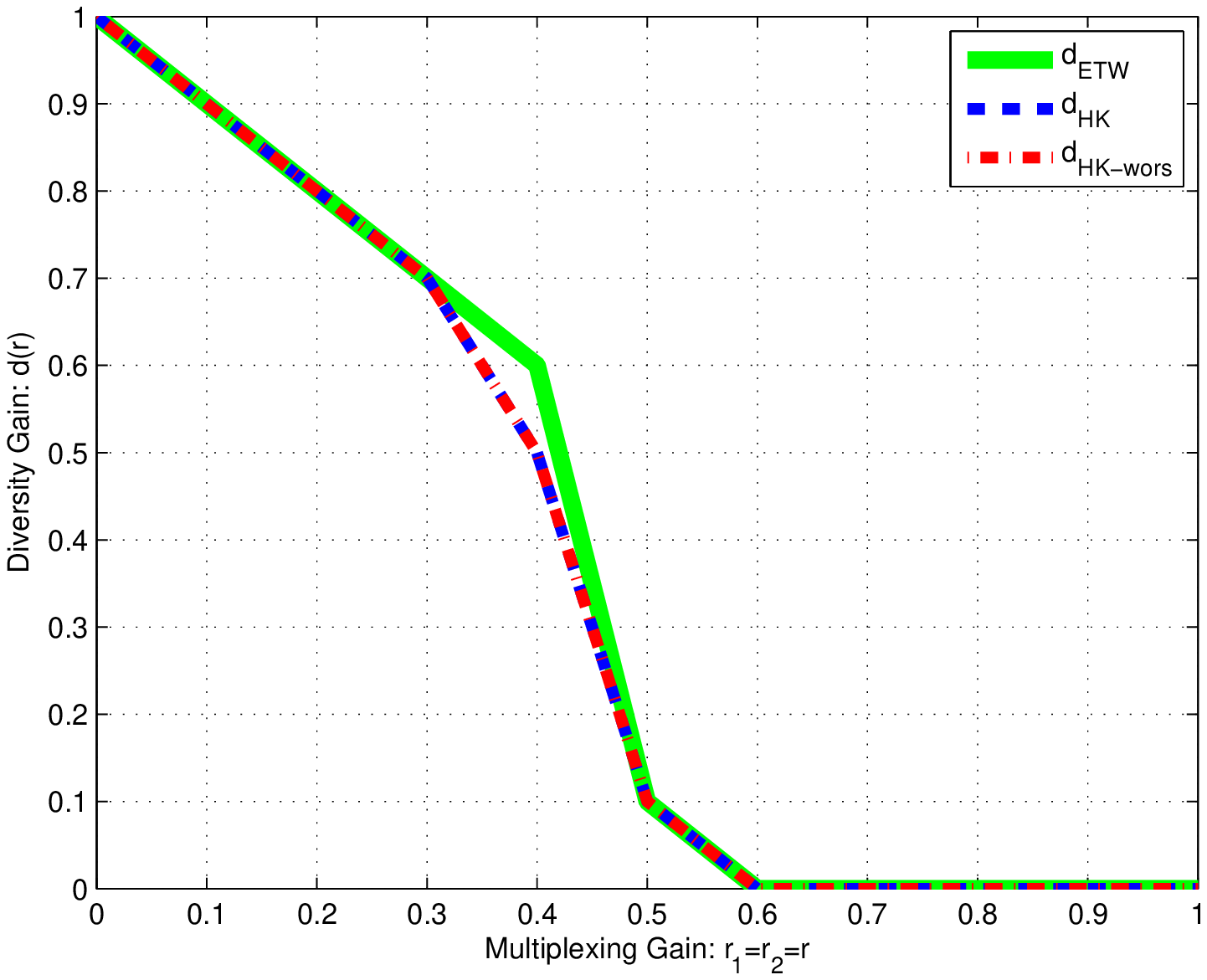}
        \caption{Symmetric channel in strong interference: diversity vs. $r_1=r_2=r$ for $\beta_{11}=\beta_{22}=1, \beta_{12}=\beta_{21}=1.1$.}
        \label{fig:sumrate1.1}
        \end{center}
    \end{figure}

    \begin{figure}
        \begin{center}
        \includegraphics[width=12cm]{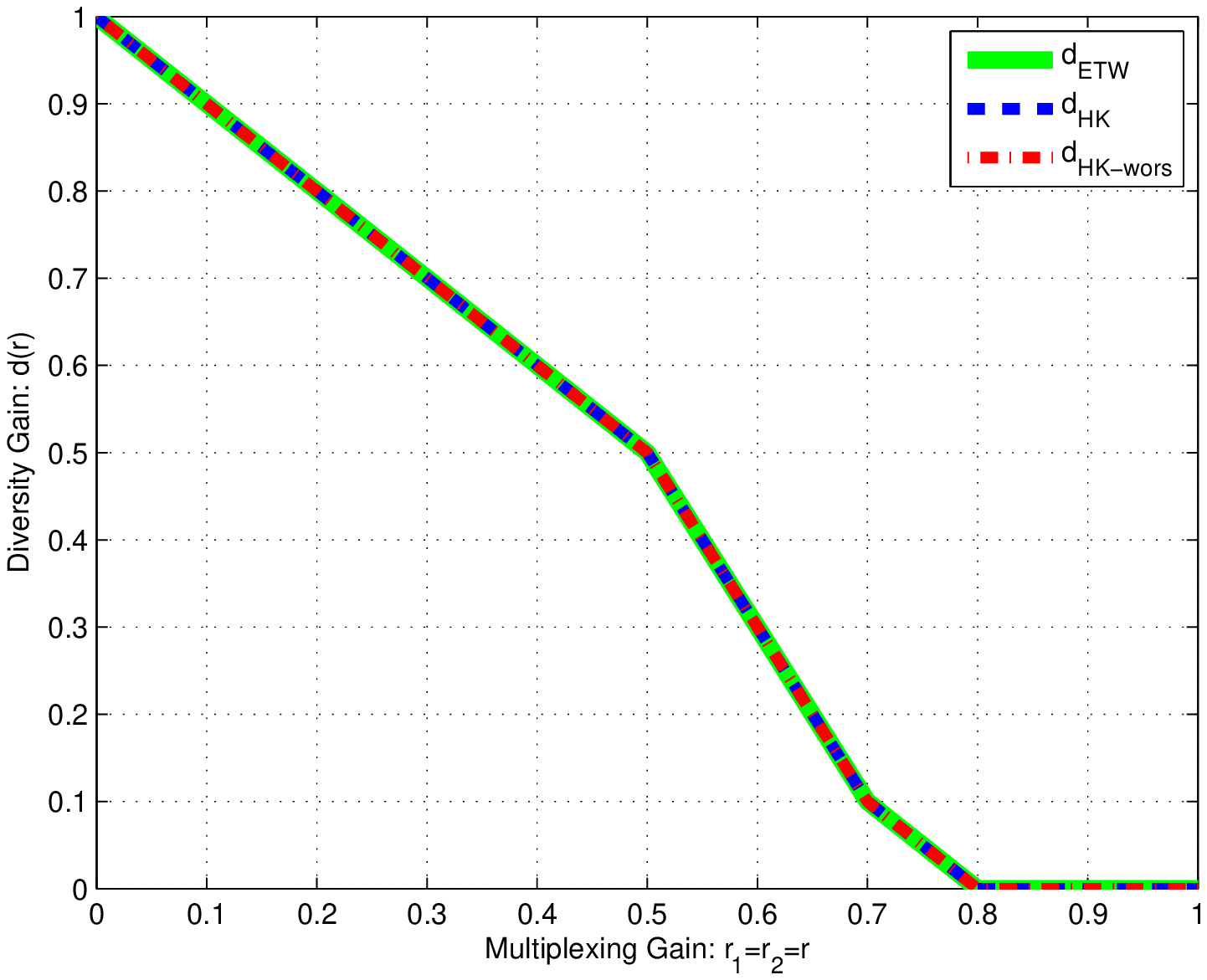}
        \caption{Symmetric channel in strong interference: diversity vs. $r_1=r_2=r$ for $\beta_{11}=\beta_{22}=1, \beta_{12}=\beta_{21}=1.5$.}
        \label{fig:sumrate1.5}
        \end{center}
    \end{figure}

We first consider symmetric channels. We set the average received power of the
direct links to $\beta_{11}=\beta_{22}=1$ and the average received power of the
cross links to $\beta_{12}=\beta_{21}=\beta\geq 0$.
In Figs.~\ref{fig:sumrate0.2},~\ref{fig:sumrate0.5},~\ref{fig:sumrate0.6},~\ref{fig:sumrate0.66},~\ref{fig:sumrate1.1}, and~\ref{fig:sumrate1.5} we plot the diversity vs. the common multiplexing gain $r_1=r_2=r$
for $\beta=0.2$,~$0.5$,~$0.6$,~$2/3$,~$1.1$, and~$1.5$, respectively.

In Figs.~\ref{fig:sumrate0.2} and~\ref{fig:sumrate0.5} (weak interference)
$d_{\rm ETW}=d_{\rm HK}$ and the dominant constraint
at low rate is the ``single user diversity''
$d_{\reff{eq:a lim}}=d_{\reff{eq:b lim}}$, while at high
rate is the ``sum-rate diversity'' $d_{\reff{eq:e lim}}$.

In Figs.~\ref{fig:sumrate0.6} and~\ref{fig:sumrate0.66} (weak interference)
$d_{\rm ETW}=d_{\rm HK}$ and the dominant constraint
at low rate is the ``single user diversity''$d_{\reff{eq:a lim}}=d_{\reff{eq:b lim}}$, at medium rate is $d_{\reff{eq:f lim}}=d_{\reff{eq:g lim}}$ and $d_{\reff{eq:c lim}}=d_{\reff{eq:d lim}}$, while at high rate is $d_{\reff{eq:a lim}}=d_{\reff{eq:b lim}}$ again.  These figures show that the expression given
for $d_{\rm ETW}$ in~\cite{leveque:izs2008} given by $d_{\rm AL}$ is not correct,
as pointed out in a remark earlier.

In Fig.~\ref{fig:sumrate1.1} (strong interference)
$d_{\rm ETW}\not=d_{\rm HK}$ and $d_{\rm HK-wors}=d_{\rm HK}$.
The dominant constraint
at low rate is the ``single user diversity''
$d_{\reff{eq:a lim}}=d_{\reff{eq:b lim}}$,
at medium rate is the ``sum-rate diversity'' $d_{\reff{eq:c lim}}$
while at high rate is again ``single user diversity''
$d_{\reff{eq:a lim}}=d_{\reff{eq:b lim}}$.

In Fig.~\ref{fig:sumrate1.5} (strong interference)
$d_{\rm ETW}=d_{\rm HK-wors}=d_{\rm HK}$
and the dominant constraint
at low rate is the ``single user diversity''
$d_{\reff{eq:a lim}}=d_{\reff{eq:b lim}}$, while at high
rate is the ``sum-rate diversity'' $d_{\reff{eq:e lim}}$.
In this case no-rate splitting is optimal.
In~\cite{AkcabaBolcskeiisit09} it was show rate splitting
is not needed in very strong interference, that is, for
$\beta_{12}=\beta_{21}=\beta\geq 2$.  Here we show numerically
that the threshold of~2 for the average interference power can be lowered.
We found by simulation that $d_{\rm ETW}$ is not achievable for symmetric
channels for $\beta\in[0.680, 1.500]$.

\subsection{Asymmetric channels}

    \begin{figure}
        \begin{center}
        \includegraphics[width=12cm]{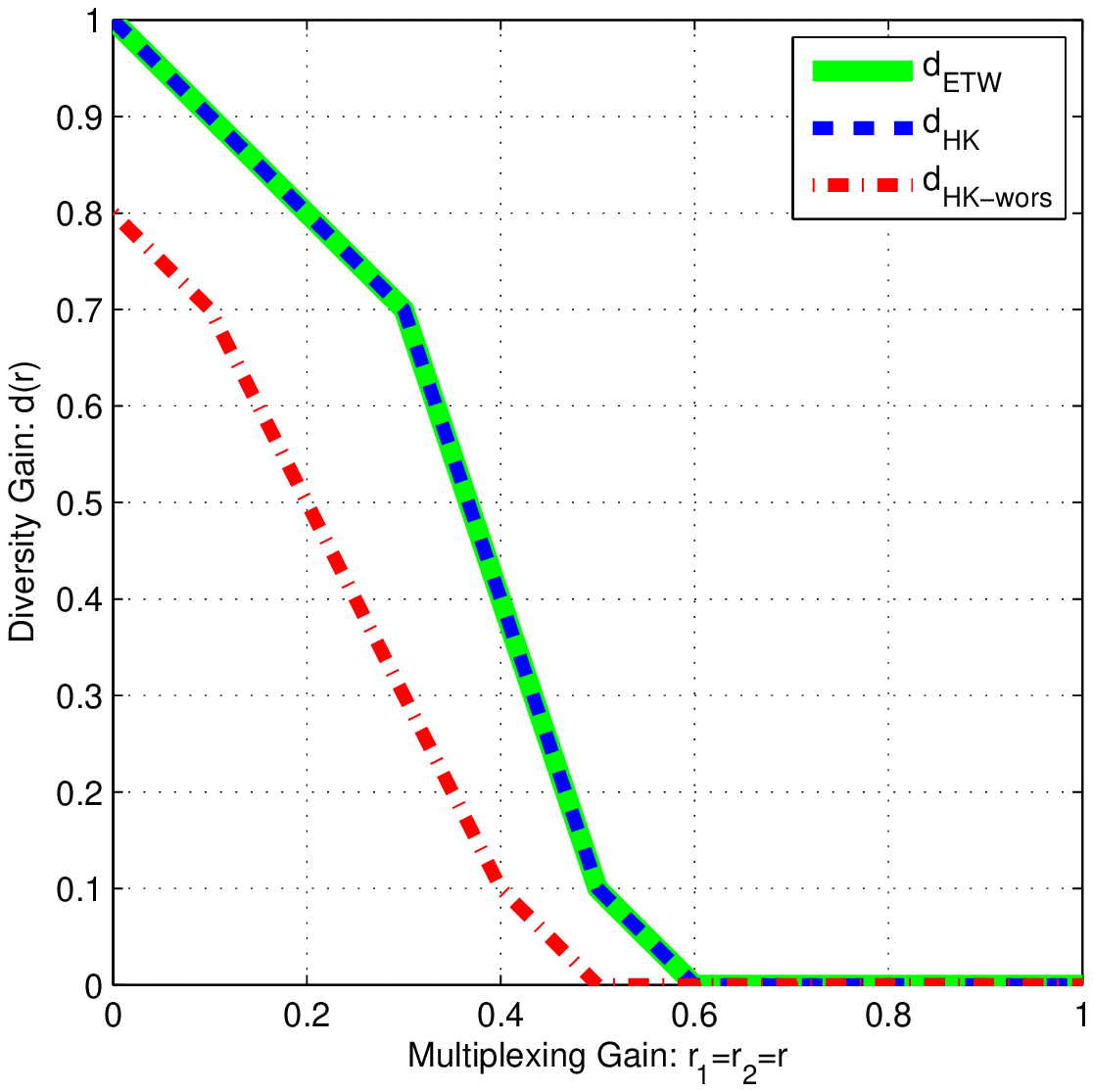}
        \caption{Asymmetric channel in weak interference: diversity vs. $r_1=r_2=r$ for $\beta_{11}=\beta_{22}=1, \beta_{12}=0.9,\beta_{21}=0.2.$}
        \label{fig:asymweak}
        \end{center}
    \end{figure}

    \begin{figure}
        \begin{center}
        \includegraphics[width=12cm]{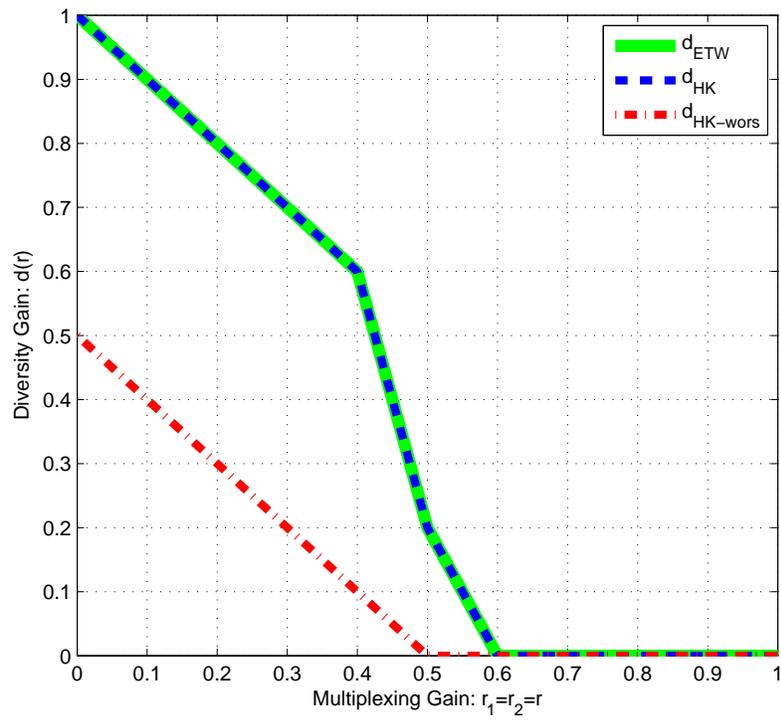}
        \caption{Asymmetric channel with mixed interference: diversity vs. $r_1=r_2=r$ for $\beta_{11}=\beta_{22}=1, \beta_{12}=1.2,\beta_{21}=0.5.$}
        \label{fig:asymmixed1}
        \end{center}
    \end{figure}


    \begin{figure}
        \begin{center}
        \includegraphics[width=12cm]{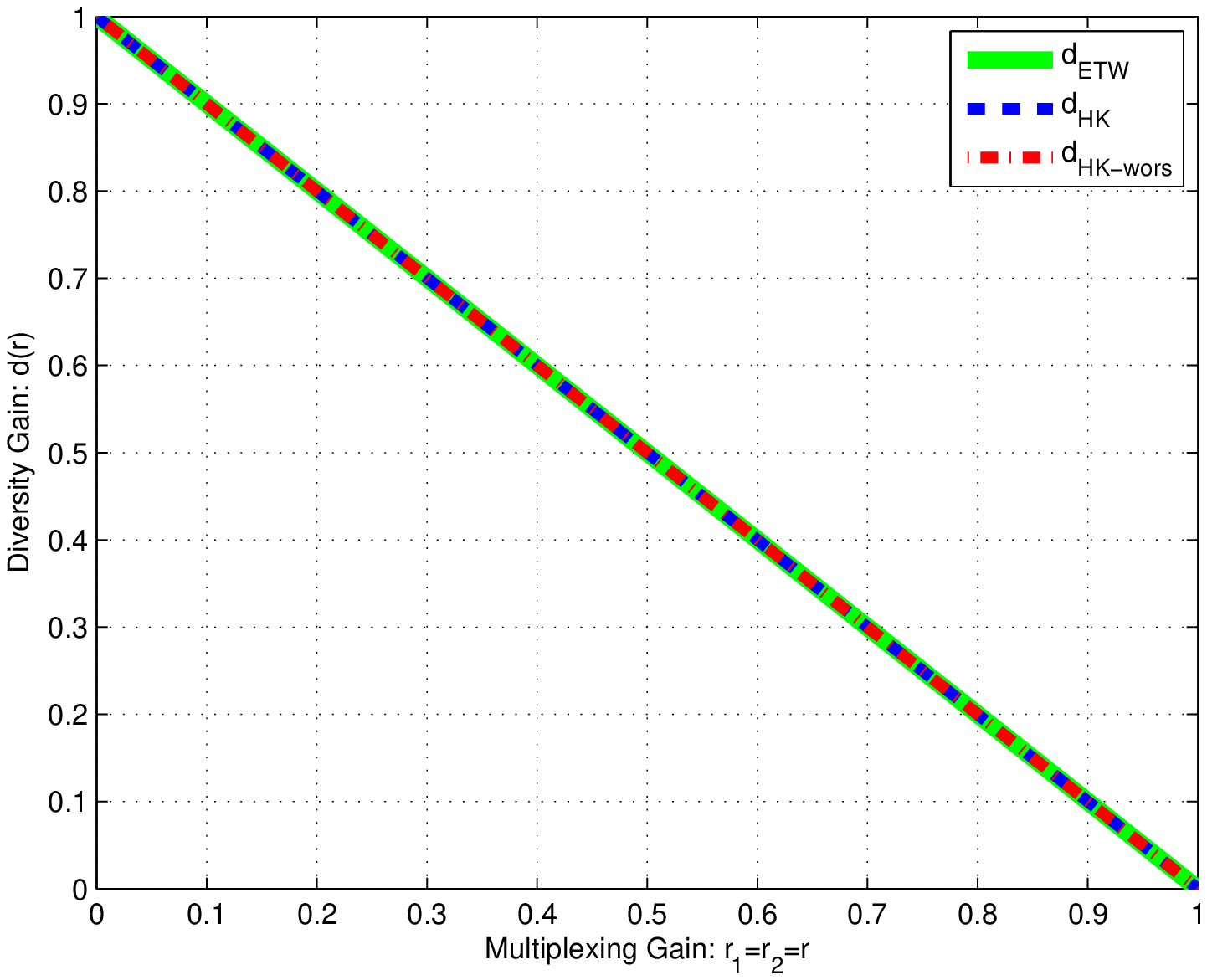}
        \caption{Asymmetric channel in strong interference: diversity vs. $r_1=r_2=r$ for $\beta_{11}=\beta_{22}=1, \beta_{12}=3,\beta_{21}=5.$}
        \label{fig:asymstrong}
        \end{center}
    \end{figure}

In Figs.~\ref{fig:asymweak},~\ref{fig:asymmixed1} and~\ref{fig:asymstrong}
we plot the diversity vs. the common multiplexing gain $r_1=r_2=r$ for asymmetric
channels with $\beta_{11}=\beta_{22}=1$ and $\beta_{12}\not=\beta_{21}$.

In Fig.~\ref{fig:asymweak} (weak interference)
$d_{\rm ETW}=d_{\rm HK}$ and the dominant constraint at low rate is the ``single user diversity'' $d_{\reff{eq:a lim}}=d_{\reff{eq:b lim}}$ since $\beta_{11}=\beta_{22}$, at medium rate is $d_{\reff{eq:f lim}}$ and $d_{\reff{eq:d lim}}$,
while at high rate is $d_{\reff{eq:a lim}}=d_{\reff{eq:b lim}}$ again.

In Fig.~\ref{fig:asymmixed1} (mixed interference)
$d_{\rm ETW}=d_{\rm HK}$ and the dominant constraint at low rate is the ``single user diversity'' $d_{\reff{eq:a lim}}=d_{\reff{eq:b lim}}$, at medium rate is $d_{\reff{eq:d lim}}$ and $d_{\reff{eq:c lim}}$, while at high rate is $d_{\reff{eq:a lim}}=d_{\reff{eq:b lim}}$ again.


In Fig.~\ref{fig:asymstrong} (very strong interference)
$d_{\rm ETW}=d_{\rm HK-wors}$ and the dominant constraint is $d_{\reff{eq:a lim}}=d_{\reff{eq:b lim}}$.
In very strong interference, i.e., $\min\{\beta_{12},\beta_{21}\}\geq \beta_{11}+\beta_{22}$,
the interference is so strong that each user can completely remove the unintended signal before decoding its own signal. In this case the
capacity region, and hence the diversity, is the cartesian product of the single user capacities without
interference.

\section{Conclusion}
\label{sec:con}
In this paper, we analyzed the diversity-multiplexing
trade-off of two-source block-fading
Gaussian interference channels without channel state information
at the transmitters. As opposed to previous works,
we considered generic asymmetric networks. We found that, a simple
inner bound based on the HK scheme with fixed power split
achieves the outer bound based on perfect channel state
information at the transmitter for wide range of channel
parameters.

\clearpage
\appendix

\subsection{Optimization of $d_{\reff{eq:HK sim c lim}}$}
\label{app:alpha fc}

The optimization problem with respect to $\alpha$ in $d_{\reff{eq:HK sim c lim}}$
involves a function of the type
\[
d_{\reff{eq:HK sim c lim}}(\alpha) =  [m-\alpha]^+ + [M-\alpha]^+ + [-m_2+\alpha]^+, \quad m\leq M.
\]
For $\alpha \leq m$
\[
d_{\reff{eq:HK sim c lim}}(\alpha) = m+M -2\alpha + [-m_2+\alpha]^+
\]
which is decreasing in $\alpha$ since, depending on the value of $m_2$,
$d_{\reff{eq:HK sim c lim}}(\alpha)$ is a straight line of slope -2 or -1.
For $m< \alpha \leq M$
\[
d_{\reff{eq:HK sim c lim}}(\alpha) = M -\alpha + [-m_2+\alpha]^+
\]
which is non-increasing in $\alpha$ since, depending on the value of $m_2$,
$f_c(\alpha)$ is  a straight line of slope -1 or 0.
For $\alpha > M$
\[
d_{\reff{eq:HK sim c lim}}(\alpha) = [-m_2+\alpha]^+
\]
which is non-decreasing in $\alpha$ since, depending on the value of $m_2$,
$d_{\reff{eq:HK sim c lim}}(\alpha)$ is  a straight line of slope 0 or +1.
Hence the function has a minimum (maybe not unique) at $\alpha = M$.

In general, it can be easily shown that the function $d_{\reff{eq:HK sim c lim}}(\alpha)$
is flat in the interval with extreme points $\max\{m,m_2\}$ and $M$. This means that
the minimum of $d_{\reff{eq:HK sim c lim}}(\alpha)$ over $\alpha\in[0,1]$ is achieved by any $\alpha$
in the interval with extreme points $\min\{1,\max\{m,m_2\}\}$ and  $\min\{1,M\}$.
In particular, we can chose $\alpha = \min\{1,M\}$.

\subsection{Optimization of $d_{\reff{eq:HK sim e lim}}$}
\label{app:alpha fe}
The optimization problem with respect to $\alpha$ in $d_{\reff{eq:HK sim e lim}}$
involves a function of the type
\[
d_{\reff{eq:HK sim e lim}}(\alpha) = [m_1-\alpha]^+ + [M_1-\alpha]^+ + [-m_2+\alpha]^+ + [-M_2+\alpha]^+
\]
with $m_1\leq M_1$ and with $m_2\leq M_2$.  With similar reasoning as
in Appendix~\ref{app:alpha fc},
it can be shown that the function is flat in the interval with extreme
points $\max\{m_1,m_2\}$ and $\min\{M_1,M_2\}$. This means that
the minimum of $d_{\reff{eq:HK sim e lim}}(\alpha)$ over $\alpha\in[0,1]$ is achieved by any $\alpha$
in the interval with extreme points $\min\{1,\max\{m_1,m_2\}\}$ and
$\min\{1,M_1,M_2\}$.
In particular, we can chose $\alpha = \min\{1,M_1,M_2\}$.
Notice that $d_{\reff{eq:HK sim e lim}}=d_{\reff{eq:HK sim c lim}}$ for $M_2=+\infty$.

\subsection{General rate splitting in the HK achievable region}
\label{app:HK gen}

The large-$x$ approximation of the HK region for a general power split
\begin{align*}
\E[|H_{cu}|^2]P_{u,private}&=   \alpha_u \, \E[|H_{cu}|^2]P_u = \frac{x^{\beta_{cu}}}{1+x^{b_u}}
      \\
\E[|H_{cu}|^2]P_{u,common}&=(1-\alpha_u)\, \E[|H_{cu}|^2]P_u = \frac{x^{\beta_{cu}+b_u}}{1+x^{b_u}},
\quad u\in\{1,2\},
\end{align*}
is
\begin{subequations}
\begin{align}
\widetilde{\mathcal{R}}_{\rm HK} =
\bigg\{
&(\gamma_{11},\gamma_{12},\gamma_{21},\gamma_{22})\in\RR^4_+:
\quad X_{ij} \defeq \beta_{ij}-\gamma_{ij}, \nonumber
\\&r_1\leq [ [X_{11}]^+ -[X_{12}-[-b_2]^+]^+ ]^+ \label{eq:HK gen a lim}
\\&r_2\leq [ [X_{22}]^+ -[X_{21}-[-b_1]^+]^+ ]^+ \label{eq:HK gen b lim}
\\&r_1+r_2 \leq
  [\max\{[X_{22}]^+, [X_{21}-[+b_1]^+]^+\}-[X_{21}-[-b_1]^+]^+]^+ \nonumber \\
&+[X_{11}-[+b_1]^+ -[X_{12}-[+b_2]^+]^+]^+
\label{eq:HK gen c lim}
\\&r_1+r_2 \leq
  [\max\{[X_{11}]^+, [X_{12}-[+b_2]^+]^+\}-[X_{12}-[-b_2]^+]^+]^+ \nonumber \\
&+[X_{22}-[+b_2]^+ -[X_{21}-[+b_1]^+]^+]^+
\label{eq:HK gen d lim}
\\&r_1+r_2 \leq
  [\max\{X_{11}-[-b_1]^+, X_{12}-[+b_2]^+\} - [X_{12}-[-b_2]^+]^+ ]^+ \nonumber \\
&+[\max\{X_{22}-[-b_2]^+, X_{21}-[+b_1]^+\} - [X_{21}-[-b_1]^+]^+ ]^+
\label{eq:HK gen e lim}
\\&2r_1+r_2 \leq
  [\max\{X_{11}, X_{12}-[+b_2]^+\}-[X_{12}-[-b_2]^+]^+]^+\nonumber\\
&+[X_{11}-[-b_1]^+ -[X_{12}-[-b_2]^+ ]^+]^+\nonumber\\
&+[\max\{X_{22}-[-b_2]^+,X_{21}-[+b_1]^+\}-[X_{21}-[-b_1]^+ ]^+]^+
\label{eq:HK gen f lim}
\\&r_1+2r_2 \leq
  [\max\{X_{22},X_{21}-[+b_1]^+\}-[X_{21}-[-b_1]^+]^+]^+\nonumber\\
&+[X_{22}-[-b_2]^+ -[X_{21}-[-b_1]^+]^+]^+\nonumber\\
&+[\max\{X_{11}-[-b_1]^+,X_{12}-[+b_2]^+\}-[X_{12}-[-b_2]^+]^+]^+
\label{eq:HK gen g lim}
\end{align}
\label{eq:HK gen lim}
\end{subequations}

The evaluation of the diversity
can be carried out similarly to the evaluation of the diversity upper
and lower bounds as done previously. In particular:

{\bf For~\reff{eq:HK gen a lim} (and similarly for~\reff{eq:HK gen b lim}):}
is as ``treating the interference as noise'' but
with interference level with $[\beta_{12}-[+b_2]^+]^+$
instead of $\beta_{12}$, that is,
\[
d_{\reff{eq:HK gen a lim}}
 =  [\beta_{11}-[\beta_{12}-[+b_2]^+]^+-r_1]^+
\in \big[[\beta_{11}-\beta_{12}-r_1]^+,\,[\beta_{11}-r_1]^+\big]
\]
The minimum value of $d_{\reff{eq:HK gen a lim}}$ is attained for $b_2\leq 0$, while
the maximum value is attained for $b_2\geq \beta_{12}$. Recall, $b_2= \beta_{12}$ is
the power split we chose in the main section of this paper.

Before the derivation of $d_{\reff{eq:HK gen c lim}}\defeq d_c$
(and similarly for $d_{\reff{eq:HK gen d lim}}$), $d_{\reff{eq:HK gen e lim}}\defeq d_e$
and $d_{\reff{eq:HK gen f lim}}\defeq d_f$
(and similarly for $d_{\reff{eq:HK gen g lim}}$) corresponding to
the constraint in~\reff{eq:HK gen c lim},~\reff{eq:HK gen e lim},~\reff{eq:HK gen f lim} let
\begin{align*}
A&=[\beta_{11}]^+, \quad B=[\beta_{11}-[b_1]^+]^+, \\
C&=[\beta_{22}]^+, \quad D=[\beta_{22}-[b_2]^+]^+, \\
E&=[\beta_{12} -[-b_2]^+]^+, \quad F=[\beta_{12}- [b_2]^+]^+,\\
G&=[\beta_{21} -[-b_1]^+]^+, \quad H=[\beta_{21}- [b_1]^+]^+,\\
r_s&=[r_1]^++[r_2]^+, \quad r_f=2[r_1]^++[r_2]^+, \quad r_g=[r_1]+2[r_2]^+.
\end{align*}

{\bf For~\reff{eq:HK gen c lim} (and similarly for~\reff{eq:HK gen d lim})
we need to solve:}
\begin{align*}
d_c &=
\min\{\gamma_{11}+\gamma_{21}+\gamma_{12}+\gamma_{22}\}
\\ &\textrm{subj. to}
  \max\{C-\gamma_{22}-[H-\gamma_{21}]^+, [G-\gamma_{21}]^+-[H-\gamma_{21}]^+\}^++[B-\gamma_{11} -[F-\gamma_{12}]^+]^+\leq r_s.
\end{align*}
We divide the optimization into two steps.
First we solve
\begin{align*}
d_{c1} &=
\min\{\gamma_{11}+\gamma_{21}+\gamma_{12}+\gamma_{22}\}
\\ &\textrm{subj. to}
 \max\{C-\gamma_{22}-[H-\gamma_{21}]^+, [G-\gamma_{21}]^+-[H-\gamma_{21}]^+\}^+\leq r_{s1},
\end{align*}
and
\begin{align*}
d_{c2} &=
\min\{\gamma_{11}+\gamma_{21}+\gamma_{12}+\gamma_{22}\}
\\ &\textrm{subj. to}
[B-\gamma_{11} -[F-\gamma_{12}]^+]^+\leq r_{s2}.
\end{align*}
Then we solve
\begin{align*}
d_{c} &= \min\{d_{c1}+d_{c2}\}
\\ &\textrm{subj. to}
\,\, r_{s1}+r_{s2}=r_{s}\defeq r_1+r_2.
\end{align*}

The optimization problem has the following three forms:
\begin{enumerate}

\item CASE 1: $G\geq H$ and $r_s> G-[H]^+$.

\begin{enumerate}

\item If $r_{s1}\leq G-[H]^+ $, then:
\begin{align*}
d_c=[C-r_{s1}]^++[G-r_{s1}]^++[B-[F]^+-r_{s2}]^+.
\end{align*}


\item  If $\min\{G-[H]^+, r_s\}\leq \min\{C,G\}$, then:
\begin{align*}
   \min d_c=\min\{(C-r_{s1})+(G-r_{s1})+[B-[F]^+-r_{s2}]^+\}.
\end{align*}

Increasing one unit of $r_{s1}$ results in two units decrease of the
object function, while increase one unit of $r_{s2}$ results in one
unit (or less) decrease of the object function. Thus
$r_{s1}=\min\{{G-[H]^+,r_s}\}$, and $r_{s2}=0$ and
\[
d_c = [C-G]^++[B-[F]^+.
\]

\item  If $\min\{G-[H]^+, r_s\}> \min\{C,G\}$, then:
\begin{align*}
   d_{c1}=\min\{[C-r_{s1}]^++[G-r_{s1}]^++[B-[F]^+-r_{s2}]^+\}
\end{align*}
Reasoning as before, $r_{s1}$ should be as large as
$\min\{C,G\}$, thus let's assume $r_{s1}=\min\{C,G\}+r_{s1}'$, then
the object function turns out to be:
\begin{align*}
   d_{c1}=\min\{[\max\{C,G\}-\min\{C,G\}-r_{s1}']^++[B-[F]^+-r_{s2}]^+\}
\end{align*}
trivially solved as:
\[
d_{c1}= [\max\{C,G\}-\min\{C,G\}+[B-[F]^+]^+-r_s']^+.
\]
Hence, by defining $T_1=\min\{C,G,G-[H]^+,r_s\}^+$, and
$T_2=r_s-T_1$ we obtain
\begin{align*}
d_{c1}=\big[C+G-2T_1+[B-F]^+-T_2\big]^+.
\end{align*}

\item If $r_{s1}\geq G-H $, then
\begin{align*}
d_{c2}&=\min [C-H-r_{s1}]^+ +[B-F-r_{s2}]^+\\
&=[[C-H]+[B-F]^+-r_s]^+
\end{align*}
and hence
\[
 d_c=\min\{d_{c1},d_{c2}\}.
\]

\end{enumerate}

\item CASE 2: $G\geq H$ and $r_s< G-[H]^+$.

If $r_{s1}\leq G-H $, by applying the reasoning in
the previous subsection and by define $T_1=\min\{C,G,G-H,r_s\}^+$, and $T_2=r_s-T_1$ we
obtain
\begin{align*}
d_{c}=\big[C+G-2T_1+[B-F]^+-T_2\big]^+.
\end{align*}

\item CASE 3: $G< H$.

In this case
\[
d_{c}=  [C-H-r_{s1}]^++[B-F-r_{s2}]^+
\]
and we trivially have
\begin{align*}
\min d_c=[C-H+[B-F-r_s].
\end{align*}
\end{enumerate}

{\bf For~\reff{eq:HK gen e lim} we need to solve:}
\begin{align*}
d_e &=
\min\{\gamma_{11}+\gamma_{21}+\gamma_{12}+\gamma_{22}\}
\\ &\textrm{subj. to}
\,\, \max\{B-\gamma_{11}-[F-\gamma_{12}]^+,E-\gamma_{12}-[F-\gamma_{12}]^+\}^+
\\&\qquad +\max\{D-\gamma_{22}-[H-\gamma_{21}]^+,G-\gamma_{21}- [H-\gamma_{21}]^+\}^+\leq r_s.
\end{align*}
We divide the optimization into two steps.
First we solve
\begin{align*}
d_{e1} &=
\min\{\gamma_{11}+\gamma_{21}+\gamma_{12}+\gamma_{22}\}
\\ &\textrm{subj. to}
\max\{D-\gamma_{22}-[H-\gamma_{21}]^+,G-\gamma_{21}-[H-\gamma_{21}]^+\}^+\leq r_{s1},
\end{align*}
and
\begin{align*}
d_{e2} &=
\min\{\gamma_{11}+\gamma_{21}+\gamma_{12}+\gamma_{22}\}
\\ &\textrm{subj. to}
\max\{B-\gamma_{11}-[F-\gamma_{12}]^+,E-\gamma_{12}-[F-\gamma_{12}]^+\}^+\leq r_{s2}.
\end{align*}
Then we solve
\begin{align*}
d_{e} &= \min\{d_{e1}+d_{e2}\}
\\ &\textrm{subj. to}
\,\, r_{s1}+r_{s2}=r_{s}.
\end{align*}

The optimization problem has the following four forms:
\begin{enumerate}

\item CASE~1: $G\geq H$ and $E\geq F$.

\begin{enumerate}
\item If $r_{s1}\geq G-H$ and $r_{s2}\geq E-F$, which requires
\[
r_s\geq [G-H]^++[E-F]^+,
\]
then
\[
d_{e1}=[[D-H]^++[B-F]^+-r_s].
\]

\item If $r_{s1}\geq G-H$ and $r_{s2}<E-F$, which require $r_s\geq [G-H]^+$, then
\[
d_{e2}=[D-H-r_{s1}]^++[B-r_{s2}]^++[E-r_{s2}]^+
\]
and
\begin{align*}
\min d_{e2}=[B+E-2T_1+[D-H]^+-T_2]^+
\end{align*}
where
\begin{align*}
T_1=\min\{B,E,E-F,r_s\}\\
T_2=r_s-T_1\\
\end{align*}

\item If $r_{s1}<G-H$ and $r_{s2}\geq E-F$, which requires $r_s\geq [E-F]^+$, then
\[
d_{e3}=[D-r_{s1}]^++[G-r_{s1}]^++[B-F-r_{s2}]^+
\]
and
\[
d_{e3}=[D+G-2T_1+[B-F]^+-T_2]
\]
where
\begin{align*}
T_1&=\min\{D,G, [G-H]^+,r_s\} \\
T_2&=r_s-T_1
\end{align*}

\item If $r_{s1}<G-H$ and $r_{s2}<E-F$, which requires $r_s< [G-H]^++[E-F]^+$,
then
\[
d_{e4}=[D+G-2T_1+B+E-2T_2-T3]^+
\]
where
\begin{align*}
T_1&=\min\{D,G,r_s,G-H\}\\
T_2&=\min\{B,E,r_s-T_1,E-F\}\\
T_3&=r_s-T_1-T_2
\end{align*}
\end{enumerate}

For the four cases above, we have
\[
d_e=\min\{d_{e1}, d_{e2}, d_{e3}, d_{e4}\}.
\]

\item CASE~2: $G\geq H$ and $E<F$.

\begin{enumerate}
\item If $r_{s1}\geq G-H$, which requires $r_s\geq G-H$, then
\[
d_{e1}=[D-H-r_{s1}]^++[B-F-r_{s2}]^+
\]
and
\begin{align*}
\min d_{e1}=[[D-H]^++[B-F]^+-r_s]^+
\end{align*}

\item  If $r_{s1}< G-H$
than
\[
d_{e2}=[D-r_{s1}]^++[G-r_{s1}]^++[B-F-r_{s2}]^+
\]
and
\[
d_{e2}=[D+G-2T_1+[B-F]^+-T_2]^+
\]
where
\begin{align*}
T_1&=\min\{D,G,G-H,r_s\}\\
T_2&=r_f-T_1
\end{align*}
\end{enumerate}

For these two cases we have
\[
d_{e}=\max\{d_{e1},d_{e2}\}
\]

\item CASE~3 $G< H$ and $E\geq F$:

\begin{enumerate}
\item If  $r_{s2}\geq E-F$, which requires $r_s\geq E-F$, then
\[
d_{e1}=[D-H-r_{s1}]^++[B-F-r_{s2}]^+
\]
and
\begin{align*}
\min d_{e1}=[[D-H]^++[B-F]^+-r_s]^+
\end{align*}

\item If $r_{s2}< E-F$
than
\[
d_{e2}=[D-H-r_{s1}]^++[B-r_{s2}]^++[E-r_{s2}]^+
\]
and
\[
d_{e2}=[B+E-2T_1+[D-H]^+-T_2]^+
\]
where
\begin{align*}
T_1&=\min\{B,E,E-F,r_s\}\\
T_2&=r_f-T_1
\end{align*}
\end{enumerate}

For these two cases we have
\[
d_{e}=\max\{d_{e1},d_{e2}\}
\]

\item CASE~4 $G<H$ and $E<F$.

We have simply:
\[
d_e=[[D-H]^++[B-F]^+-r_s]^+.
\]

\end{enumerate}

{\bf For~\reff{eq:HK gen f lim}:}

We need to solve:
\begin{align*}
d_f &=
\min\{\gamma_{11}+\gamma_{21}+\gamma_{12}+\gamma_{22}\}
\\ &\textrm{subj. to}
\,\, \max\{A-\gamma_{11}-[F-\gamma_{12}]^+,E-\gamma_{12}-[F-\gamma_{12}]^+\}^++[B-\gamma_{22}-[F-\gamma_{12}]^+]^+\\
&\quad\quad
    +\max\{D-\gamma_{22}-[H-\gamma_{21}]^+,G-\gamma_{21}-[H-\gamma_{21}]^+\}^+
\leq r_f.
\end{align*}
We divide the optimization into two steps.
First we solve
\begin{align*}
d_{f1} &=
\min\{\gamma_{11}+\gamma_{21}+\gamma_{12}+\gamma_{22}\}
\\ &\textrm{subj. to}
\,\, \max\{A-\gamma_{11}-[F-\gamma_{12}]^+,E-\gamma_{12}-[F-\gamma_{12}]^+\}^++[B-\gamma_{22}-[F-\gamma_{12}]^+]^+\leq r_{f1},
\end{align*}
and
\begin{align*}
d_{f2} &=
\min\{\gamma_{11}+\gamma_{21}+\gamma_{12}+\gamma_{22}\}
\\ &\textrm{subj. to}
\,\, \max\{D-\gamma_{22}-[H-\gamma_{21}]^+,G-\gamma_{21}-[H-\gamma_{21}]^+\}^+\leq r_{f2}.
\end{align*}
Then we solve
\begin{align*}
d_{f} &= \min\{d_{f1}+d_{f2}\}
\\ &\textrm{subj. to}
\,\, r_{f1}+r_{f2}=r_{f}.
\end{align*}

The optimization problem has the following four forms:

\begin{enumerate}
\item CASE~1: $E\geq F$ and $G\geq H$:

\begin{enumerate}
\item
$r_{f1}\geq E-F$ and $r_{f2}\geq G-H$, which requires $r_f\geq [E-F]^+ -[G-H]^+$.

If $r_{f1}'\leq \max(A',B')-\min(A',B')$, then
\[
d_f=d_{f1}+d_{f2}=[\max\{A',B'\}-r_{f1}']+[D-H-r_{f2}]^+
\]
and
\[
\min d_{f1}=[\max(A',B')+[D-H]^+-r_f]^+
\]
with
\[
A' = \big[A-F-[E-F]^+\big]^+, \quad
B' = [B-F]^+
\]

If $r_{f1}'> \max(A',B')-\min(A',B')$, then
\[
\min d_{f2}=\frac{[A'+B'-r_{f1}']^+}{2}+[D-H-r_{f2}]
\]
where
\begin{align*}
r_{f2}&=[D-H-R_f]^+\\
r_{f1}&=r_f-r_{f2}\\
r_{f1}'&=[r_{f1}-[B-F]^+]^+
\end{align*}
and
\[
d_f=\max\{d_{f1},d_{f2}\}
\]

\item
If $r_{f1}\geq E-F$ and $r_{f2}<G-H$, which requires
$r_f\geq E-F$.

If $r_{f1}'\leq \max(A',B')-\min(A',B')$, then
\[
d_f=[\max\{A',B'\}-R_{f1}']+[D-r_{f2}]^++[G-r_{f2}]^+
\]
and
\[
\min d_{f1}=[\max(A',B')+D+G-2T_1-T_2]^+
\]
where
\begin{align*}
T_1&=\min(D,G,G-H,r_s)\\
T_2&=r_s-T_1
\end{align*}

If $r_{f1}'\geq \max(A',B')-\min(A',B')$, then
\[
\min d_{f2}=\frac{[A'+B'-r_{f1}']^+}{2}+[D-H-r_{f2}]
\]
where
\begin{align*}
r_{f2}=\min\{\max\{D,G\},r_f\}^+\\
r_{f1}=r_f-r_{f2}\\
r_{f1}'=[r_{f1}-[B-F]^+]^+
\end{align*}
and
\[
d_f=\max\{d_{f1},d_{f2}\}
\]

\item
If $r_{f1}<E-F$ and $r_{f2}\geq G-H$, which requires
$r_f\geq G-H$, then
\[
d_f=[E-r_{f1}]^++[A-r_{f1}]^++[B-F]^++[D-H-r_{f2}]^++
\]
thus
\[
d_f=[E+A-2T_1+[D-H]^+-T_2+[B-F]^+]^+
\]
where
\begin{align*}
T_1&=\min\{E,A,E-F,[r_f-[B-F]^+]^+\}\\
T_2&=r_f-[B-F]^+-T_1
\end{align*}

\item
If $r_{f1}<E-F$ and $r_{f2}< G-H$, which requires
$r_f<[E-F]^++[G-H]^+$, then
\[
d_f=[E-r_{f1}]^++[A-r_{f1}]^++[B-F]^++[D-r_{f2}]^++[G-r_{f2}]^+
\]

Thus
\[
\min d_f = [E+A-2T_1+D+G-2T_2+[B-F]^+-T_3]^+
\]
where
\begin{align*}
T_1&=\min\{E,A,E-F,[r_f-[B-F]^+]^+\}\\
T_2&=\min\{D,G,G-H,[r_f-T_1-[B-F]^+]^+\}\\
T_3&=[r_f-T_1-T_2]^+
\end{align*}
and hence
\[
d_f=\min\{d_{f1},d_{f2},d_{f3},d_{f4}\}
\]
\end{enumerate}

\item CASE~2: $E\geq F$ and $G<H$:

\begin{enumerate}
\item
$r_{f1}\geq E-F$ which requires $r_{f1}\geq E-F$.

\[
d_{f1}=[[[A-F]^+]^++[B-F]^+-r_{f1}]^++[D-H-r_{f2}]^+
\]
and
\[
\min d_f=[[[A-F]^+]^++[B-F]^++[D-H]^+-r_f]^+
\]

\item
$r_{f1}<E-F$.

\[
d_{f2}=[E-r_{f1}]^++[A-r_{f1}]^++[B-F]^++[D-H-r_{f2}]^+
\]
and
\[
\min d_{f2}=[[E+A-2T_1+[D-H]^+]^+-T_2+[B-F]^+]^+
\]

\[
d_f=\max\{d_{f1},d_{f2}\}
\]
\end{enumerate}

\item CASE~3: $E< F$ and $G\geq H$.

\begin{enumerate}
\item If $r_{f2}\geq G-H$, which requires $r_{f2}\geq G-H$,
then
\[
d_{f1}=[[[A-F]^+]^++[B-F]^+-r_{f1}]^++[D-H-r_{f2}]^+
\]
and
\[
\min d_f=[[A-F]^++[B-F]^++[D-H]^+-r_f]^+
\]

\item If $r_{f2}<G-H$
then
\[
d_{f2}=[[A-F]^+]^++[B-F]^+-r_{f1}]^++[D-r_{f2}]^++[G-r_{f2}]^+
\]
and
\[
\min d_{f2}=[[D+G-2T_1+[A-F]^++[B-F]^+-T_2]^+
\]
where
\[
T_1=\min\{D,G,G-H,r_s\}\\
T_2=r_s-T_1
\]
and hence
\[
d_f=\max\{d_{f1},d_{f2}\}
\]
\end{enumerate}

\item CASE~4: $E< F$ and $G< H$.

We simply have
\[
d_f=[[A-F]^++[B-F]^+-r_{f1}]^++[D-H-r_{f2}]^+
\]
and thus
\[
\min d_f=[[A-F]^++[B-F]^++[D-H]^+-r_f]^+.
\]

\end{enumerate}

\bibliographystyle{unsrt}
\bibliography{ywRef}

\end{document}